# QUANTUM THEORY OF ECONOMICS

## V.I. ZVEREV* and A.M. TISHIN

**Physics Department of M. V. Lomonosov Moscow State University, Leninskie Gory, Moscow**

**119992**

## Abstract

In the given work the first attempt to generalize quantum uncertainty relation on macro objects is made. Business company as one of the economic process participants was chosen by the authors for this purpose. The analogies between quantum micro objects and the structures which from the first sight do not have anything in common with physics are given. The proof of generalized uncertainty relation is produced. With the help of generalized uncertainty relation the authors wanted to elaborate a new non-traditional approach to the description of companies' business activity and their developing and try to formulate some advice for them. Thus, our work makes the base of quantum theory of economics.

* Corresponding author: vi.zverev@physics.msu.ru

# Preface

«Life always refashions and turns itself; it is much higher than our stupid theories».

Boris Pasternak «Doctor Zhivago»

# Introduction

Two seemingly unconnected factors were the reason to begin our research. The first one is that according to the statements of world leading macroeconomists there is NO appropriate economic theory able to describe all processes in the world economy nowadays. [1] For example, on March 2, 2004 Alan Greenspan, then the head of Federal Reserve System, in his speech 'Current Account' in New York Economical Club said '…as far as I know despite the analysts' attempts there is no any model that can predict the dynamic of the rates of exchange better that it can be done by tossing a coin. I know about thousands of people who try to predict it and some are quite successful. Very much like winners of pennies matching'. We of course do not want to say that world leading economists still have not understood the essence of obvious indeterminacy in any economical activity. Any person who is far enough from both physics and economics understands rather well that it is absolutely impossible to predict for example the cost of any product or some service in some time. It is that thing that is considered as indeterminacy in economy in our ordinary life.

The first known to us attempt to describe this phenomenon from scientific point of view dates from the 60-s when a book 'Physics and economics' by Kuznetsov was published. Nowadays this book is considered to be a rare book. The author tries to prove the groundlessness of the term distinctness speaking about economics. He only mentions the existence of indeterminacy in economical processes but he does not say a word about nature of this phenomenon and of course does not give any numerical estimations of this indeterminacy. In the present work on the base of the detailed analysis of quantum and economical objects we are going to be the first ones who demonstrate the opportunity of quantum approaches application to indeterminacy description. We examine such an economical object as a company. We must say that our work is different from the one that is devoted to determined chaos and practically does not have anything in common with it.

Let us make this statement clearer. The phenomenon of determined or dynamic chaos is a complex unpredictable behavior of determined *nonlinear* system. This theory states that simple systems which consist of little quantity of elements and which obey *determined* rules (without elements of chance) can sometimes show accidental, rather complex and unpredictable behavior.

This suddenness is of principal and irremovable nature. Such a phenomenon is considered to be a chaos.

In our work we are going to speak about opposite things and try to describe systems with undetermined rules which explain unpredictable behavior of these systems. Since the systems to be described are linear we can use the quantum approach to their description. It is well known that quantum mechanics describes only linear systems.

Superposition principle is one of the proofs of quantum economical systems linearity. It is known that it is quantum mechanics where superposition principle ‘celebrates victory’ because it is the theory where there is a finite probability for the system in one state to be in another state at the same time. It is that fact that is called a superposition principle in quantum theory. The linearity of fundamental theory in this branch of physics is the reason of superposition principle appearance. It is obvious that this principle is fulfilled for business-structures. A good example is some conglomerated company that can successfully operate in different non-connected parts of the market.

The second one is that the determined limits of quantum theory applicability still are not defined. It was Niels Bohr who first paid attention to these borders. [2] He drew the conclusion that it is probably possible to extrapolate uncertainty relation on macro objects, i.e. it was the statement that an observer probably has influence not only on the micro objects but on the macro objects as well ‘Continuous metabolism between an organism and the environment is necessary for life, that is why it is impossible to separate the organism as physical-chemical system from the environment. Therefore we can think that any attempt to find that precise border which will allow to accomplish full physical-chemical analysis will lead to the changes in metabolism incompatible with the life of the organism.

Despite that Bohr did not determine the borders between microscopic quantum system and the observer with macroscopic device he made a plain statement about the principal difference between the theory of quantum objects based on Schrödinger equation and the classical theory in which Schrödinger equation can not be applied. It is necessary to mark that we should not connect concepts of quantum and classical objects with geometrical size of these objects. According to Bohr’s statement such a connection reflects only historic conditions of quantum mechanics origin during the analysis of phenomena in microscopic physical systems. [3] Nowadays quantum phenomena with macroscopic size are known and the Universe itself is often considered as the united quantum object.

For example, it is known through the experiments that there are carbon atoms in the Galaxy with n~1000 (n is main quantum number), atomic radius r~0.1 mm and wavelength of transition between two excited levels λ~18 м [4]. Hence, the discussed border does not have an objective character and exists only in the physical model which we use to describe microscopic world.

Quantum mechanics restores the idea of the united world and mutual connection of all phenomena. This idea was often discounted in classical physics. Sharp distinctions between waves and corpuscles, between objects for observing and medium already do not exist. Interconversions of matter are brought to the forefront. We should agree with very accurate statement of David Bohm ‘Apparently we should reject the idea that the Universe could be divided on separate parts and apply a submission that the whole world is united. In every case where quantum phenomena play crucial role we will find out that separate ‘parts‘ of the Universe can essentially change in time because of inevitable and inseparable connections between them. Thus we lead to the picture of the Universe as the picture of indivisible but flexible and constantly changing object. ‘ [5]

Here we speak about so called inseparability (integrity) of quantum system, i.e. about indissoluble connections between any part of the system and the system itself and about impossibility to divide the system onto parts even in theory. Behavior and properties of separate parts are defined by the whole system, in other words these properties correlate. It was reasoning about inseparability of quantum system and quantum correlations that refuted any attempts to dispute fullness of quantum theory. These attempts began from the famous EPR paradox and ended with the disproof of hidden variables theory. In this case we think that the most natural example of this principle fulfillment is any business-structure. We can hardly examine work of concrete employees ignoring the activity of the whole company and even work of the concrete company in isolation from the activity of the other players in any market segment and from state agencies of regulation. Thus the system which describes any business-company and the results of its activity is quite inseparable. The existence of correlations among its parts which submit to quantum laws will be noted below.

Taking into consideration all said above we should try to answer the following question. Is it possible to introduce the analogue of uncertainty relation for separate elements of macro systems, i.e. for elements with very small effective mass (in comparison with the effective mass of macro system) and on very small space intervals (definition of the effective mass will be introduced later)? Is it possible, for example, to show that according to uncertainty relation there is no ‘trajectory‘ of the company in macro-economics as it is impossible to predict the precise trajectory of the particle in quantum mechanics? For example who could predict the ‘trajectory‘ of it might seem ‘eternal‘ Fannie Mae and Freddie Mac? The artificial correction of their ‘diving’ trajectory by US government saved them from the unavoidable collapse. It is a pity but we can not say the same about Lehman Brothers which had been accomplishing financial activity fro 158 years and burst like a soap-bubble on September 14, 2008. How much is it justified to extend strict laws of quantum theory to financial activity of business-structures? We hope that our work is a search for firm principles which determine «physics» of companies. Laws of organized human performance and business are such a kind of performance without any doubt must remain unchangeable as well as laws of physics. Since our work is counted on the wide circle of readers among whom there can be people very well acquainted with economics and not very well acquainted with quantum mechanics during our work we will give supplement and to some extent reductive interpretation of the bases of quantum theory.

We can summarize all said above. Since all the environment is considered to be quantum in frames of modern quantum mechanics and it can be considered continuous only in classical approximation some objects from this environment in special conditions can show quantum behavior.

It is worth mentioning that a number of physical terms are already widely used in economical theory. For example, inertia, braking, acceleration, velocity, energy, fragility (fragility theory), elasticity of demand and supply etc. In the book 'Managing Corporate Lifecycles' by I. Adizes we can fins such purely physical terms as critical mass, energy, impulse, inertia, force which as the author believes are instinctively clear to the reader. [6]

# About measuring and prediction problems

## About measuring problem in quantum mechanics

The main feature of classical approach to phenomena description is supposition of full independence of physical processes from observational conditions. In classical physics we suppose that it is always possible to 'spy on' the phenomenon without having an influence on it. Frankly speaking if we 'spy on' the process from different points of view (describe it in different reference frames) forms of the process will be different. But dependence of process passing form on reference frame moving has been always taken into consideration. This aim can be achieved by simple converting of coordinates of one reference frame to coordinates of another one. At the same time changing of the character of passing process which happens during transition from one reference frame to another does not change the process itself at all. That is why in classical physics we are able to talk about independence of the phenomenon itself on means of observing.

Quantum mechanics has shown that an opposite thing exists in the world of micro objects, where the possibility of observing itself supposes the existence of definite physical conditions which can be connected with the essence and character of passing of the phenomenon observed Setting of these conditions is not only indication of reference frame used but it also demands their detailed characteristics. In other words the speech is about role of ‘measuring instrument‘ and procedure of measuring in quantum mechanics. It is clear that the procedure of any physical quantity measuring requires bringing the measuring device in the system itself. The procedure of measuring is a result of interaction between the object examined and the instrument of examination. We got used to thinking that the reverse influence of the device on the object is always lost in the noise. It is impossible to neglect such an influence in quantum theory. The presence of the measuring device in quantum system changes the system and makes it change in the way different from the one without the instrument. The simplest example of this fact is hypothetic attempt of electron’s location finding. In ordinary ‘classical‘ life when we want to know if there is a necessary book on the table we usually look at the table and find the book on it or not. It is the end of ‘classical‘ searching process. And if we want to know the location of the electron on the table we will have to do the same deeds. But the fact is that light beam which carried the information about the electron to the retina of the eye and then to the brain was reflected by the electron some tiny time ago. It was enough for the electron and light quantum to interact. As the result of such an interaction the electron could easily change its location on the table or even ‘fall down‘ from the table. More complex measuring which is held with the help of special apparatus undoubtedly changes the system to a greater extent.

The neglect of this fact in classical physics is abstraction which can be called overemphasizing of physical process. If we apply this concept we are able to consider any physical process as happening in itself independently of the fact if it is in principle possible to observe it or not.

Common sense and everyday observations prompt to us that there is a negligible part of events happening themselves in real life and especially in business-world. When we want to know how intensively water in a pan with closed cover boils we raise the cover and after that the water will inevitably boil not so intensively. As is easy to see we speak even not about measuring but only about observation. In a similar manner no employee, no human being, no department and eventually no company exists isolated. For example, preparing for company’s IPO or the attempt of an investor to ‘measure‘ the company from profit point of view make the company demonstrate fast future growth and gain rising to the investor. Also such attempts make successful companies search for innovative ways of development, more active attraction of clients, interaction with partners, and struggle against competitors, in other words the whole macro system is changed. Often executive management of the company or its stockholders want to have more information about employees’ or different subdivisions’ of the company work buy getting supplement data from ‘inside‘. So it is natural that management control or any other ‘measuring of physical quantity‘ is possible only with inner state of the company changing, with any external influence otherwise ‘measuring‘ is impossible.

## Prediction character of quantum mechanics

All said above points to the fact that quantum mechanics must be based on such ideas about motion which are principally different from ideas of classical physics. To formulate appropriate statements we should firstly find out character of target settings in quantum mechanics. Typical target setting in quantum theory is prediction of next measuring result according to known results of previous measurings. Essential differ between quantum and classical theories is in the fact that description of quantum system state is realized by smaller quantity of variables, i.e. quantum description is less detailed than classical one. [7]

So we have very important consequence regarding predictions of quantum mechanics. Whereas classical description is enough to predict motion of mechanical system precisely in future, less detailed quantum description is obviously not enough for this. The main aim of quantum mechanics is in finding probability of getting this or that result of measuring in any concrete experiment. These probabilities together with average value of the dynamic variable known as the result of infinite quantity of experiments are the results of theory. Quantum theory can not answer what precise value we will get but it can answer with what probability we will get this value.

Let us make more exact what is probability in quantum mechanics and in our everyday life. As for quantum theory probability in it is understood according to the strict scientific definition from probability theory. In this case probability is some numerical characteristic of every elementary outcome. In other words it is some function of distribution which shows how often this or that event happens in the series of experiments. All outcomes in probability theory are considered to be accidental. The problem is the following. For the outcome to be considered as accidental it must satisfy four demands. They are indeterminacy; all multitudes of outcomes are known; the experiment can be repeated infinite quantity of times. This demand gives birth to such an important notion in probability theory as frequency of outcome. Frequency equals the ratio of number of realizations of concrete outcome and general number of repeated experiment. The last demand is statistical stability of frequency.

And when all these demands are satisfied in quantum theory simply because of natural properties of quantum objects and introduction of probability is completely proved things become worse in everyday life and in the economy. At least we can speak about indeterminacy in company or market behavior or shares cost as the result of accident or planned events in economy, politics, and society with confidence (recent examples are General Motors Corporation, United Airlines, Mechel). Although it is obvious that people who were creators of well-organized financial collapses (e.g. financial crisis in 1997-1998) knew the probability of what would happen better than ordinary people. First of all exterior observers or analysts of course do not know precisely what will be the behavior of executive management in a new situation and moreover sometimes even executive officers themselves do not know precisely what to do especially if they did not meet such a problem before. And it is clear that it is absolutely impossible to predict all multitudes of probable outcomes for the company or to organize a series of experiments during which to choose the best variant of behavior. We are simply not able to carry out the controllable iterative experiment in which all variables excepting one can be fixed and get different results changing only that variable we are interested in. Titus Lucretius Carus said ‘Everything run, everything changes. It is impossible to enter the same water two times. ‘ [8] That is why there no any sense to speak for example about the probability of oil will cost 150 $/barrel if military invasion of Iran begins or not. The invasion will begin or will not begin, oil will cost 150 $ or will not cost. It strangely enough does not mean that probability of any of these events equals one half. Let us remember a joke about the probability to meet a dinosaur in the street ‘What is the probability to meet a dinosaur in the street? Obviously, fifty-fifty. Why? Well, we will meet it or we will not meet.‘ It approves that fact that in real life market reaction to any event is not only undefined but even the probability of this or that model of behavior can not be known since probability does have any meaning as used here. But nevertheless there is a number of consulting firms, analytical bureaus and also training courses such as ‘It is simple to play on exchange‘, ‘I will teach you to earn money‘ whose aim is essentially determination of market reaction to any external influence. It is interesting to note that in their activity they use the same instruments as probability theory namely the analysis of this or that segment of market statistics during the long period of time. We should declare that some of them are successful in their business. Therefore it is possible to say that sometimes such a simple and ordinary understanding of probability sometimes gives us much more accurate and very often much more useful results than strict understanding of probability in quantum mechanics. And even if the fans of mathematical formalism don't like it we present our apologies.

Thus we can draw the intermediate conclusion that even qualitative analysis i.e. narrow-minded view or life experience allows us to insist that some results of quantum theory can be applied for description of business-structure.

## Comparative analysis of typical properties of quantum micro objects and business-companies.

Quantum mechanics as independent science was born in 1900 with Max Planck idea of light quantum. From the very beginning it was appealed to explain a number of physical phenomena which could not be explained in the frames of classical physics. Subsequently with the development of experimental technique real quantum objects which exist in nature were discovered. Their behavior could not absolutely be understood without quantum theory. The most typical quantum objects are elementary particles. What should be the properties of the object to be able to use quantum laws for its behavior description? In other words what is typical quantum object?

But we should make a reservation that pursuing our object (see Abstract) we content ourselves with only those properties of quantum objects analogues of which can be found in business-companies. As the work is devoted to absolutely new scientific direction the present level of knowledge still does not allow to find the analogues of ALL properties of particles in economical theory.

So let us enumerate the properties of an object which can be considered to state surely that the object is quantum.

### a) Typical masses

It is known that mass of electron is m=9, $1*10^{-31}$ kg, mass of proton equals1836m, neutron – 1839m, heavy electron – 207m. So for our purpose we can take ~ $10^{-30}$ kg as mass of the electron. Typical mass of classical object is ~ $10^{2}$ kg. Classical object in our work is any object around us in our everyday life. We should make more exact the fact that we do not give limit values of classical objects masses here but take some average value calculated from the estimate of classical masses objects around us. Frankly speaking different bulk materials as a rule save their physical properties peculiar to bulk state until the size of 100 nm what means the masses equal $10^{-17}$ – $10^{-18}$ kg. So from the first sight the dispersion between quantum and classical physics is in the range of 13 and 32 orders. Indeed this dispersion is greater because the description of macro molecules also obeys quantum laws but the masses of these molecules can reach $10^{-21}$ kg so the approximate dispersion is between 9 and 32 orders. As the limit size of macro molecules is not known we can suppose that the lower boarder (9 orders) can transform soon in 6 or even 3 orders. As it was said above precise boarders of quantum mechanics application are also still unknown. So we can say that even from the mass point of view these bounds are also not determined. As we know uncertainty relation can be applies to the phenomena ‘with the particles of a very small mass in very small space intervals’.[7]

And what happens in economics? But before we name typical orders we will definite an effective mass. We will explain a little bit later why we have chosen namely business-companies for our research but we should make a reservation that any participant of economical process is a good example of quantum indeterminacy manifestation in economical theory. We offer to introduce a mass like physics do, i.e. a mass is a measure of company inertia. It shows how difficult for a company to change or simply turn the vector of its trajectory (development direction) while the effective mass increases. The change of profile of a giant company with a huge amount of employees, wide net of branches and departments all over the world gives birth to much more serious difficulties than the same activity of a small family business. It is well known how difficult it was for the hew head of Kimberly-Clark Darwin to sell all pulp and paper mills and to put up money in Huggies and Kleenex brands. One more example. In the early 1970-s Kroger and A&P were old companies and their assets were put up in a system of usual provision shops. Kroger decided that stores with huge amount of departments are their future and that is why it began to close, remake or replace the stores which were behind the times. The period between 1970 and 1990 was a difficult time of reforms but by the early 1990-s Kroger has rebuilt all the system and has become the largest network of grocery stores in the USA. At the same time the majority of A&P stores were from the 1950-s.

By the effective mass we will understand the number of employees and clients of the given company, in other words all participants of economical process organized by the given economical object. The fact is that taking clients into account is quite reasonable. Any company works with clients straight or in implication. Retailing companies, food networks, shops work with clients straight and struggle for them. Industrial and high-tech companies supply their production through middlemen but it does not prevent to have information about the amount of buyers and according to carry out one or another policy in dependence on this number increasing or decreasing. For the given company the quantity of clients is very important at that each client should be taken into account with weight 1 since we speak about the company whose client he is. He is free to be the client of another company which does not need to have anything in common with the company we want to calculate the effective mass of. That is why the effective mass includes the total quantity of clients and employees. Moreover, according to special relativity theory an object with nonzero mass already possesses nonzero energy (rest energy). If we take only employees into account we get the value of rest energy but it is obvious that normally working business-company possesses nonzero kinetic energy while moving in business-space (interaction with clients). Economical objects with relatively small mass will be the objects of our research because the effective mass of the most giant company is small in comparison with population of the Earth which is more than 6.5 bln people nowadays or $10^{10}$ **unified economical** **mass units** (**ue**). Chemists and physicists some 50 years ago introduced the notion of atomic mass unit (amu) as 1/12 of $^{12}C$ mass ($m_C$=1.995 $10^{-26}$ kg, $m_C$ /12=1.66 $10^{-27}$ kg) and a notion of moth a little bit earlier. Like physicists the economists can define **unified economical mass unit** (**ue**) and economical **moth. 1 ue** can be defined as 1 employee + 1 client of the company. It is worth saying that there are companies that are permanently changing their clients (e.g. duty free) but however the average quantity of the clients is quite constant and even in this case we can calculate the effective mass using formula «1+1».

We should say that we do not answer the question about the critical value of the company mass. By this value we will understand that quantity of employees and clients which leads to the company collapse or its transformation into classical object. Now we can only say that although the formula 1+1 is universe for the companies from different branches of world economy the critical value depends on market segment and external conditions. Lehman Brothers just before the bankruptcy had a mass of 130 thousand ue (30 thousand clients + 100 thousand employees) while Citigroup has a mass of 200, 3 mln. ue (200 mln. clients + 300 thousand employees) and still has not collapsed. We should mention that there are a number of companies fro which we can speak about the quality of the effective mass. For example, General Electric aims at getting more profit from industrial but not financial direction (in ratio 60% to 40%). We understand that for such giant companies we are not quite right to speak about small effective mass (paying attention to the approximate borders of quantum theory of economics application) but as it will be shown below even these quasi-classical objects obey quantum laws let alone small and middle companies with effective masses of 1000. It is possible for the companies with more than 100 mln. clients to take employees into account as some amendment of the first order since the dispersion of 3 orders is essential. This approach can be also applied to the specific participants of economical process such as cultural and art workers. An artist (in general) and his assistants (about 10 people) can make a product that will be bought by 1 mlrd. people. However the formula 1+1 is correct.

Let us introduce the dispersion of effective masses. The mass may equal units of mass i.e. only several people are the participants of economical process (e.g. a watchmaker and a couple of his clients). It means the mass is ~ $10^0$ ue. On the other hand a number of employees and clients of an average company ~ $10^3$ ue while for the large companies it can be $10^7$-$10^8$ ue. Thus, we can say that the value of the effective mass can be 2-10 orders less than $10^{10}$ ue what is compared with the values for physical quantum systems. So taking these values of effective masses into account we can state that examine the behavior of the companies with small effective masses.

**b) Typical times**

In this paragraph we should mention such a property of quantum objects as instability. To be honest enough some objects do not possess this property. All elementary particles excepting photon, proton and neutrino are unstable. It means that they turn into other particles spontaneously and without any external influence. For example neutron spontaneously breaks up into proton, electron and electron antineutrino. It is impossible to predict the concrete time of concrete neutron decay; each decay is accidental and depends on huge quantity of factors. The analogous situation takes place as regarding birth of particles. Both these processes are purely probabilistic and statistical phenomena. Do not we see the same in business community where large companies let alone small ones because of the competition (interaction with competitors) can disappear from the market or change the type of business activity or can find themselves under state control? As the result of concluding bargains or negotiations a large company can be broken to small pieces (if it is unstable) and each ‘piece‘ will be absorbed by the larger and more successful company. Thus the decay of the company because of its *instability* happens. Instability in its turn is determined by the huge number of presuppositions and all of them can not be taken into consideration (e.g. ineffective management, unfavorable state of the market, pressure of state bodies of regulation etc.)  The same thing happens in case of new companies’ birth when the part of administrative stuff leaves the company. The same factors which led to the decay of one company can successfully contribute to fusion of two or more companies and the birth of more successful, more *stable* company. [9] Just as there is only a few stable particles in nature (photon, proton, electron) there can be only some stable companies in the frames of separate macro economical system. The problem of particles and companies stability is quite disputable. For example it is known that life time of ‘stable’ proton is $10^{31}$ s. Last events in the American market prove that we can hardly find any stable company. It is possible that Merrill Lynch (founded 1914) is going to share Lehman Brothers’ fate soon. Basing on the results of our work we can suppose that if the effective mass of the company essentially increases in some market it is not a quantum object anymore and its behavior aims at determinism. Most likely this transfer from quantum to classical behavior happens in some range of effective mass values. In this case a company does not feel the quantum character of the environment anymore and should be broken onto new quantum objects with less effective mass of it will need ‘collar and lead’ in the form of state control to be able to ‘walk’ in this environment. So we can draw the conclusion that those companies whose effective mass essentially increases will meet serious difficulties soon and some of them will simply disappear. And if one of ‘built to last’ (according to Collins’s definition) is among them… well - it wasn’t fortunate enough. That day when Lehman Brothers collapsed Alan Greenspan said that ‘we should get used to living in the world where seemingly eternal companies collapse.' And the events which now happen with largest world companies surprisingly looks like the birth and annihilation of quantum objects. Readers who are especially interested in the aspect of so called 'eternal' can take a closer look at Jim Collins's book 'Built to Last' [10]. In general the history of stock company (and all 'eternal' companies are stock ones) is not longer than 200 years. Among world wide known companies arms company Beretta is the oldest. It was founded in 1526. As for the champions in the list of long-livers they are small business companies which as a rule are not very well known. The oldest company in the world nowadays is Japan building company Kongo Gumi, which was founded 1430 years ago. [11]

In the context of problems of stability and instability discussion we can not help mentioning such a notion as life time of quantum micro object. Every unstable elementary particle is characterized by its own life time. The less is life time, the more is the probability of decay of the particle. For example, life time of heavy electron is 2, $2*10^{-6}$ c, hyperons - about $10*^{-10}$ c. In the 70-s of XX century 100 particles with tiny life time$10^{-22}$ – $10^{-23}$ c were discovered. They were name delta-resonances. Typical life time of classical objects around us is in average ~ $10^{1}$ years, i.e. about $10^{9}$ c. Thus the dispersion is 32 orders (like in case with masses). There is such a notion as 'a firm that lives one day' in the economy. Juridical life time of such firms lays in the range from one quarter to several years. Average life time of usual companies (we do not speak now about stable companies from the group of 'Built to Last' which can be called quasi-stable) is about ~ $10^{1}$ years, i.e. about $10^{3}$ days. The dispersion is 3 orders (like in case with masses). Coincidence of the orders of dispersions put the following idea into our mind. The laws of quantum indeterminacy can be applied in economical theory.

### c) Tracks of decay and interconversions

It is notable that quantum objects can have different tracks of decay. For example positively charged A-meson can break up into heavy electron and its neutrino or neutral A-meson, positron and electron neutrino. For the given particle we can predict neither time of decay nor track of decay. There are a huge number of legal and illegal tracks of companies' 'decay' in the business world. Finally all of them lead to the disappearance of organized business-structure. Among these variants are bankruptcies, breaking up of the company (conversion) and also so called raider captures (absorption of small firms by large ones). Nobody of course can predict the track of company 'decay' until this decay happens. But it is probably more correctly to speak not about the 'decay' or 'death' of the company but about interconversions of companies. Indeed even the implacable enemies of the company who would like to destroy it nevertheless want to have maximum profit from the organized economical process especially when this process was not organized by them. The same thing is in nuclear physics where the decay of the particle is not decay in its first meaning. The decay of the particle is an act of interconversion of one particle to other particles. The best proof of this fact is probability for one particle to have several tracks of decay. For example it is not right to say that Russian RAO EES (Russian Joint-stock Company United Energetic Systems) consisted of the multitude of small companies before its conversion on July 1, 2008. The quantity and quality of new companies appeared after conversion depended on huge amount of factors but everything happened as it happened. Everyday experience teaches us that taking an object to pieces means to know what this object consists of. The idea of such an analysis (the idea of breaking up) reflects the key features of *classical* mind. While the transition to micro objects this idea partly works e.g. a molecule consists of atoms, an atom consist of the nucleus and electron, the nucleus consists of protons and neutrons. However it is the end of breaking up idea because breaking up of proton or neutron does not show any inner structure if these particles. So we can not state that the decay of quantum micro object means that this object consists of decay products. Another example. Magnetic molecules $Mn_{12}$ and $Fe_8$ which consist of metallic and organic parts will not have total spin S=10 after removing some atoms from the organic cover although the value of total spin is provided by the ferromagnetic structure of 12 atoms of Mn and 8 atoms of Fe. As for company structures we see the same thing. All said above approves the thesis about inseparability of quantum economical system once more.

### d) Ground and excited states of quantum system

Instability is peculiar to practically all quantum objects. For example radioactivity (spontaneous transformation of one chemical element's isotopes to the isotopes of another element accompanied by emanation of particles) shows that atomic nuclei also can be instable. Atoms and molecules in excited states are also instable. They spontaneously turn into ground or less excited state. Companies and even whole macro-economics also aim to reach ground unexcited state in their activity if the transfer was not connected with structural or technological changes or was not supported (like in recent history with Fannie Mae and Freddie Mac) by the state or any other external injections. Otherwise a new state becomes instable and the company returns to the ground state incurring losses. A good illustration of this principle is General Motors activity for the last five years. Having received a supplement portion of energy as decreasing of pressure of taxation the company became the leader in the sales market. But it became the leader because of supplement but not cumulative energy and that is why GM incurs losses for several years while returning to the ground state or to the one of close to it meta-stable states. Because of increasing of salary and equipment costs the company without government support will inevitably return to it. We can’t help mentioning a notion of the separate country or region stability. The size of the given country is not so important. The main thing for the system is its quantum character. If the excited state of this country economy is based on the energy swapping from less developed countries or regions the tiniest mistake in this process let alone its stop will lead to the greater instability of economy and the attempts of politicians including military attempts to support such a state. Here we speak not only about oil a gas. The swapping of country’s economy can be made by means of financial and other sources.

### e) Spin

Spin is one of the most important peculiar features of the micro object. Spin is own moment of impulse of the micro object. Its presence is connected neither with motion of the object nor with external conditions. Probably it is very difficult and unclear definition for the reader without physical education but modern microscopes allow to know the precise spin direction of the surface magnetic atom. Spin should be considered as innate property of quantum micro object as a mass or an electric charge. Presence of supplement degree of freedom of the object is the most important consequence of spin existence. The number of these supplement degrees of freedom is the number of possible spin projections on the chosen direction. At this point of quantum theory of economics development we can not say precisely what indicator of business activity can be identified with spin. But the fact that the company always has several variants of behavior among other players in the market, the possibility of maneuver speaks well of the presence of some property of the company which is analogous to spin. According to this we can produce an interesting observation. It is known that information about spin allows to judge about the behavior of an object among the same objects. In other words we can get information about statistical properties of the micro object. It is well known that there are only two groups of elementary particles in nature according to their statistical properties. They are particles with integer spin and particles with half-integer spin. Micro objects from the first group (integer spin) are able ‘to occupy‘ the same energy state without any limits on the amount of the particles in this state. We say that such particles obey Bose-Einstein statistics. Micro objects from the second group are able ‘to occupy‘ one energy state only being there alone. If the given state is ‘occupied‘ no other micro object can occupy it. Such particles obey Fermi-Dirac statistics. It is connected with properties of wave function symmetry which in their turn come from identity principle for particles. Bluntly speaking we can not number quantum objects, then transpose them and say that we get a new combination of them. Quantum objects are principally undistinguishable! Obviously we can not say the same about companies. Each of them is unique.

Every company which a little bit respects itself has it own logotype nowadays. We can agree that the structure of modern companies is very much the same. Every company has executive management, stuff of employees, similar system of departments and subdivisions. At last there are the same printers, scanners, Xeroxes in every office. From this point of view we can characterize any company and probably we will not see the difference among hundred companies. The same situation is with atoms. One atom has definite quantity of protons, neutrons and electrons, another one has also definite but another quantity of these particles in its structure. Nevertheless, despite the fact of analogous description we say that the first atom is oxygen and another is hydrogen. It is that diversity of these characteristics that provides us with Periodic Table of chemical elements. All companies have the same but essentially different characteristics. Even if one twin brother works in one company and another brother is in another company these companies can be situated in different places, there can be different phone' numbers etc. What is more despite they are twins the brothers think in a different way as regarding the companies' policy. Therefore, business companies are principally distinguishable. But they strangely enough obey the same mentioned above quantum statistics depending on the size and the area of activity. Giant transcontinental corporations were made for business in conditions of almost free competition in almost ideal market. Moreover a powerful stimulus for this competition presence is antimonopoly law which exists in every country with market economics. The aim of this law in those countries where it really works is to make the companies stay in quantum state i.e. make them have small effective mass in the given market. That is why such companies having some part of the market feel comfortably in their niche among the competitors. To our mind they very much look like bosons. On the other hand a small undertaker-monopolist is a typical Fermi particle which forces out aliens and prevents young businessmen from starting their business. Such a situation was typical in Russia in the 90-s of the XX century. Thus we have shown that quantum economical objects rather satisfactorily obey Bose-Einstein and Fermi-Dirac statistics even without identity principle for them.

### f) Fundamental interactions

The problem of interaction between describable structures is one of the most crucial in our work. What leads to changing of their motion state? Now we can only suppose that quantum economical systems interact by means of virtual or real particles which are quanta of appropriate fields and carriers of appropriate interactions (similar to photons, bosons, gluons). Since physicists nowadays work on unification of interactions in the frames of quantum field theory it is quite possible that companies interact by means of some new type of quanta which is not still discovered. It is a fact that these interactions are not investigated and not discovered in scientific literature but in the early 1930-s physicists also knew only gravitational and electromagnetic interactions with the help of which they could not explain the complex structure of atomic nuclei whereas X-ray and electron were discovered already in 1895-1897. In addition, the fact those weak and electromagnetic interactions are only manifestations of electro-weak interaction was shown only in 1957-1967. Moreover despite the centuries-old investigations and undoubted deserts of Galileo, Newton and Einstein and also success of contemporary scientists in the development of new gravity theory nobody has not discovered gravitational waves propagating with light speed. Even the experiments of finding out the precise value of gravity constant are carried out still.

Another example. There is no theory that can fully describe high-temperature superconductivity but high-temperature superconductive materials are already produced in industry. So from our point of view the problem of the absence of information about the character of interactions and force between quantum economical objects is not a good reason not to use quantum approach to their description. Everybody knows that there are forces of 'friction', 'attraction', 'repulsion' between separate employees (e.g. gravitational interaction between 2 employees 1 meter from each other is $10^{-9}$ N). We do not offer to describe interactions in the companies by means of simple physical forces because everything is much more complex. But it is force of friction among employees and subdivisions that can lead to split in the stuff and the retirement of some employees and even to the birth of new companies.

In conclusion we would like to mention relativist effect of mass defect. Mass of a molecule, an atom or a nucleus equals sum of masses of particles this object consists of minus some quantity called mass defect. Mass defect equals the energy necessary to break the object to constituents (binding energy) divided by velocity of light. The more is binding energy, the more is mass defect. We will probably not find numerical examples of mass defects for business companies but it is clear that an inevitable decreasing of the effective mass while the process of companies' fusion happens. It happens because of possible reduction of the staff, sort of management, systematization of financial flows etc. Brazil Semco is a good example. If there were no mass defect raider would not aim to control large companies, divide them onto separate parts and sell separately. Mass defect also should be taken into account while calculating the effective mass f the company.

So the analysis of typical properties of typical quantum objects and business structures behavior is the evidence of the supposition that there are no any limitations to use quantum approach for the description of business activity. A more detailed talk about interaction between companies, spin of the company and presence or absence of antimatter at least – is a matter of future.

# Uncertainty relation and its consequences

## Uncertainty relation

Actual uncertainty of the next experiment result gives birth to a number of conceptual problems of quantum theory. The first one is explanation of physical reason of experiment's results plurality. The answer is Heisenberg uncertainty relation received by him in 1927. Uncertainty relation is a quantitative formulation of quantum object's peculiar features. The fundamental significance of this relation is the fact that with its help Heisenberg managed to show that polysemy of numerical value of dynamic variable in quantum mechanics is determined not by mistakes of every measuring but by physical properties of this variable. [5] So uncertainty relation mathematically expresses indeterminacy of quantum object motion, indeterminacy originates from change in object state as the result of measuring act. Physicists of classical traditions could not recognize introduction of indeterminacy. Einstein expressed this by the famous words that 'God does not play dice' [12] As a result of this several attempts to disprove uncertainty relation were made. Finally they failed to find appropriate examples.

Let us now produce the derivation of generalized uncertainty relation for quantum economical objects. Since it was proved above that there are no any limitations to apply quantum approach for description of business companies' activity the derivation of generalized uncertainty relation is based on mathematical formalism of quantum theory. The derivation is based on the idea from the Physical encyclopedia. [13] It is known that the state of quantum system (quantum object) is characterized by observable variables (observables). An observable in its turn is that physical quantity which value we want to know in the given experiment. Mathematical apparatus of quantum mechanics is such that every observable (every physical quantity) corresponds to Hermite operator. Spectrum of an operator (diagonal elements of its matrix) corresponds to the probabilities of different values of the physical quantity that can be measured. Two observables $A$ и $B$ are called compatible if their commutator equals zero i.e. $[A,B]=0$. According to the definition of commutator $[A,B]=AB-BA$. Let us examine two incompatible observables $A$ и $B$, connected by the following equation their commutator equals the third observable multiplied by imaginary unit):

$$[A,B]=iC\,, \tag{1}$$

Let us hermitian mate (1).

We mention that according to the definition of the Hermite operator:

$$A^{+} = A, \tag{2}$$

where upper index $^{+}$ means the operation of Hermite coupling.

We get:

$$[A,B]^{+} = (AB - BA)^{+} = B^{+}A^{+} - A^{+}B^{+} = BA - AB = -[A,B] = -iC^{+}, \tag{3}$$

So from comparison of we have

$$iC^{+} = iC = [A, B], \tag{4}$$

what means that $C$ is a real observable since according to (4) the Hermite operator corresponds to it.

Let in some state we measure values of $A$, $B$ and $C$.

According to the definition of variate's dispersion:

$$DA = \langle (A - \langle A\rangle)^{2}\rangle, \tag{5}$$

Let average value $\langle A\rangle$ is $a$, the dispersion of $A$ is $\langle \tilde{A}^{2} \rangle$,

and eventually we have:

$DA = \langle \tilde{A}^{2} \rangle$,

$DB = \langle \tilde{B}^{2} \rangle$, где $\tilde{B} = B - b$ (6)

Let us introduce an operator $\hat{F} = \tilde{A} + i\lambda\tilde{B}$ (7)

It is obviously Hermite since

$$\hat{F}^{+} = \tilde{A} - i\lambda\tilde{B}. \tag{8}$$

That is why operator $\tilde{F}$ does not correspond to any real observable.

Let us introduce an operator

$$\hat{G} = \hat{F}^{+}\hat{F}. \tag{9}$$

This operator already correspond to the real observable since

$\hat{G}^{+} = \hat{F}^{+}(\hat{F}^{+})^{+} = G$. (10)

Spectrum of $\hat{G}$ is nonnegative since

$$\begin{aligned} &\hat{G}\,|\,\gamma\rangle = g\,|\,\gamma\rangle, \\ &\langle\gamma\,|\,\hat{G}\,|\,\gamma\rangle = g\langle\gamma\,|\,\gamma\rangle, \end{aligned} \tag{11}$$

where $|\,\gamma\rangle$ is eigenvector of $\hat{G}$, which eigenvalue $g$ corresponds to.

Owing to the fact that

$$\langle\gamma\,|\,\gamma\rangle = 1 \tag{12}$$

we have

$$g =<\gamma|\hat{G}|\gamma>=<\gamma|\hat{F}^{+}\hat{F}|\gamma>=<\chi|\chi>\geq 0, \qquad (13)$$

where $<\chi|=<\gamma|\hat{F}^{+};|\chi>=\hat{F}|\gamma>$.

Axioms of John Neumann (1932) are in the base of mathematical formalism of quantum theory. The third axiom is about an average value:

The average value reproduces the properties of measured quantity, i.e. if

$$a_n \geq 0, \text{then} <\hat{A}>\geq 0, \qquad (14)$$

where $a_n$ - measuring results of *A*.

Then according to the third axiom from (13) we have $<\hat{G}>\geq 0$.

According to (7) – (9) we get quadratic inequality relative to *λ*:

$$\begin{aligned}&<\hat{G}>=<(\tilde{A}-i\lambda\tilde{B})(\tilde{A}+i\lambda\tilde{B})=<\tilde{A}^2+\lambda^2\tilde{B}^2+i\lambda(\tilde{A}\tilde{B})-i\lambda(\tilde{B}\tilde{A})>=\\&=<\tilde{A}^2+\lambda^2\tilde{B}^2+i\lambda[\tilde{A},\tilde{B}]>=<\tilde{A}^2+\lambda^2\tilde{B}^2+i\lambda iC>=<\tilde{A}^2+\lambda^2\tilde{B}^2-\lambda C>=\\&=DA+\lambda^2 DB-\lambda<C>\geq 0.\end{aligned} \qquad (15)$$

For (15) to be fulfilled discriminant of quadratic equation $\lambda^2 DB-\lambda<C>+DA=0$ must be nonpositive, i.e.

$$<C>^2-4DADB\leq 0. \qquad (16)$$

From (16) we get:

$$DADB\geq\frac{<C>^2}{4}. \qquad (17)$$

Let us extract square root from (17) and let $\sqrt{DA}=\delta A, \sqrt{DB}=\delta B,$ then (17) will be:

$$\delta A\delta B\geq\frac{<C>}{2}, \qquad (18)$$

where $\delta A, \delta B$ - root-mean-square deviations (square root from dispersion) of*A* and *B* correspondingly, and $<C>$ - quantum ensemble average of *C*.

The relation we received in frames of our suppositions completely coincides with canonical Heisenberg uncertainty relation for coordinates and impulses to the equivalent

$$A\equiv x, B\equiv p, C\equiv\hbar_{gen}.$$

(19)

We should not forget that (19) are operator equations. $\hbar_{gen}$ is a number multiplied by unitary matrix of infinite dimension.

Thus we get generalized uncertainty relation for coordinates and impulses:

$$\delta x\delta p\geq\frac{\hbar_{gen}}{2}. \qquad (20)$$

$\hbar_{gen}$ (generalized) is the analogue of Planck's constant in quantum theory of economics. Search of its numerical value is the task of separate research. In the given one-dimensional case the coordinate uncertainty in (20) describes the uncertainty in inertia centre of the company coordinate or uncertainty of the exact location if its effective mass. The fact that this relation in physics works only on the atomic level is explained by the small numerical value of Planck's constant. For the (20) to work in macroscopic scale $\hbar_{gen}$ should have the values compared with other quantities from (20).

What does (20) mean from economical point of view? For the larger clearness let us introduce business space with dimension N where the company exists No matter how abstractedly seems the introduction of multivariate space for the description of company evolution such an approach is quite reasonable. Business activity depends on a huge amount of factors (variables). They can be either endogenous (depend on the company itself e.g. staff quantity, amount of products etc.) or exogenous (correspondently not depend on the company e.g. taxes, world market situation etc.) Let us check qualitatively whether (20) fulfils in reality.

It is well known that with the help of modern electronic data bases any chief of the company can know the financial history of the company with certain accuracy. He can get data from yesterday evening or today morning or even from several hours or minutes ago. The determination of the financial state at the very moment is impossible because bank information is not received every second but it takes some time to fulfill the electronic inquiry. And we hardly find a chief who can precisely predict the financial state of the company tomorrow morning for example. It is well known that sent and approved payments sometimes are not entered in an account in time. Let alone that any stranger can publish an old information about the bankruptcy as in case of United Airlines because of which the cost if its shares has lost 76%. Thus the indeterminacy in motion of a company description is observed.

Let we want to define precisely the coordinate of the company in business space at the concrete moment of time. Let this coordinate will be the volume of money, which the company possesses now. But as it follows from the common sense such an information can be known if we stop all economical activity of the company at that moment (block bank accounts), in other words if we know the precise value of company's impulse. But for such a stop all employees and clients who form an effective mass of the company must immediately finish for the moment all their activity e.g. bargaining, conclusion contracts, negotiations, selling etc. It means that we should make an effective velocity of business activity equal null, which gets in a formula for an effective impulse:

$$p_{eff} = m_{eff} v_{eff} \qquad (21)$$

As we understand it is absolutely impossible. Thus we can not simultaneously know precise values of the coordinate and impulse of the company.

And vice versa let us suppose that we precisely know the impulse of the company at the moment, i.e. the precise quantity of employees and clients and also the number of bargains at the same moment. This is hypothetically possible but anybody hardly needs it. But in this case we will not be able to find the coordinate precisely, e.g. the same amount of money, because business activity is not stopped and value of the coordinate is changing every moment.

So we have shown that because of generalized uncertainty relation we can not absolutely precisely and simultaneously know values of the coordinate and impulse of the company in the business space.

For a quasi-stationary case we have uncertainty relation for energy and time:

$$\Delta E \Delta t \approx \hbar_{eff} \qquad (22)$$

We will understand $\Delta E$ as a quantitative measure of different forms of motion of quantum macro-objects and all types of interactions, and $\Delta t$ is time necessary for measurement (the longer we hold the pan open the less precisely we know what was in it before the lid open). The energy notion is actively used by the number of authors. In Adizes work we can read '…since each moment of time the system has energy store. This fact is known from physics. The quantity of energy can be increased in the systems that successfully interact with the environment. But every moment of time the system has fixed quantity of energy.' Business process can be characterized by the phrase 'exchange of energy (of employees) to money'.

So the economical meaning of the relation (22) is the following. If the energy of the quantum system is measured to $\Delta E$, than the time which corresponds to this measuring has the minimal uncertainty according to (22).

## Quantitative proofs of trajectory absence for business companies (quantitative consequences of generalized uncertainty relation)

In this paragraph let us consider (22) for the description of business activity in more detail. For simplicity sake and without detriment to commonness let us consider company evolution in two-dimensional space. In the light of (22) let us choose usual time as time variable (it is quite natural to observe any evolution in time). For energy variable we obviously choose such a variable that can be measured in money. Functions of money from economical point of view quite well correspond to the energy of the company expressed in money equivalent. One of eventual results of business activity is the exchange of the energy contributed to money after all. Moreover choice of money as energy variable is determined by the following fundamental fact. As it is known that classical theory deals with *continuously* changing quantities whereas in quantum theory we have to do with *discrete* processes. Quantum is indivisible portion of energy and nobody managed to carry out an experiment to discover part of quantum. So it was shown that transfer of energy process is of *discrete* character. The same situation is in the economy. The natural indivisible and minimal portion of energy is cent in the USA, kopeck in Russia, eurocent in EU etc. Thus it is quite reasonable to imagine the transfer of energy process as receiving or return of some amount of money which is divisible by the minimal portion of energy by the way. Quarterly and annual net profit of the company was chosen as a concrete energy characteristic of the business activity. It is connected only with greater reliability of statistics on net profit that we use in our work. The same thing can be said also about other characteristics of the company (e.g. annual sales volume, capitalization etc.)

Let us return for a little bit to quantum mechanics. Among different types of measuring definition of electron's coordinates plays the most important role in quantum theory. Let we measure successive values of electron's coordinates in equal terms. Generally speaking the results will not correspond to any smooth curve. On the contrary the more precisely we measure the coordinates the more stick-slip and disorderly nature of the results we see. The absence of the trajectory of the electron is the direct consequence of uncertainty relation. More or less smooth trajectory can be produced if we measure the coordinates with little degree of accuracy e.g. condensation of drops of steam in the cloud chamber. (fig.1.)

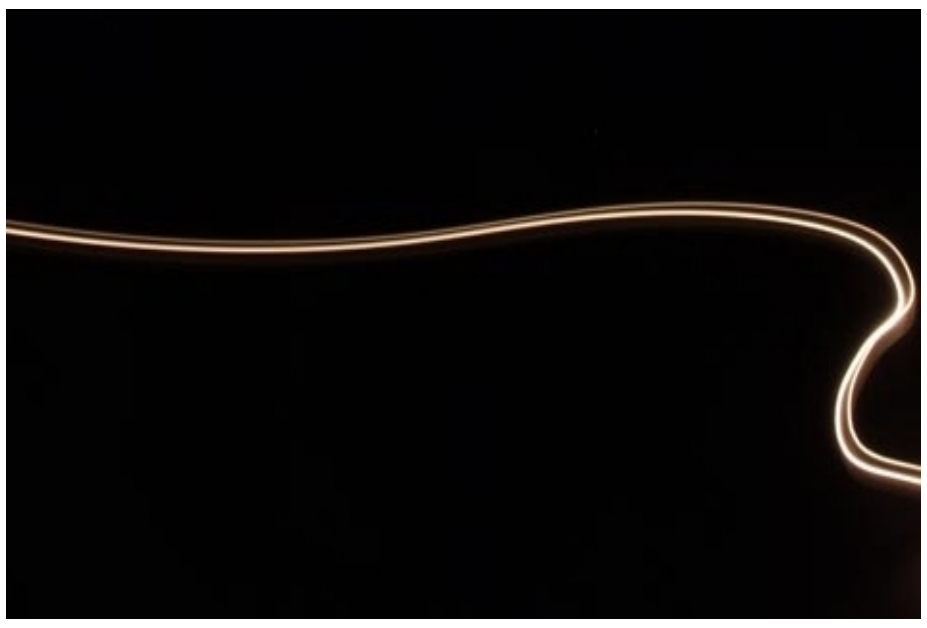

Fig.1. Electron trajectory in the cloud chamber [14]

The same picture can be produced for other elementary particles (quantum micro objects) in modern detectors such as bubble chamber, spark chamber etc. (fig. 2.)

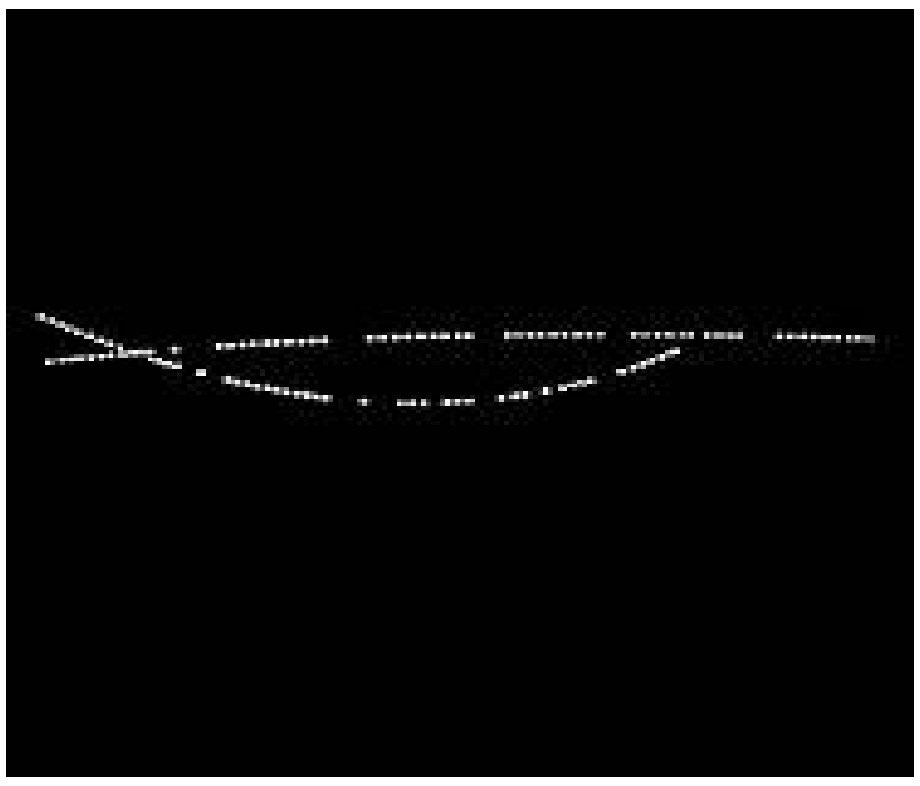

Fig.2. Particles' tracks in the bubble chamber[13]

The word trajectory is used here to underline that little degree of accuracy when we are able to speak about the trajectory of quantum micro particle.

In general the absence of trajectory is a typical property of namely quantum object.

Let us now track trajectories of world leading companies in two-dimensional business space using data about their *annual* net profit (bln USD) from 2000 to 2007. (fig. 3-5.)

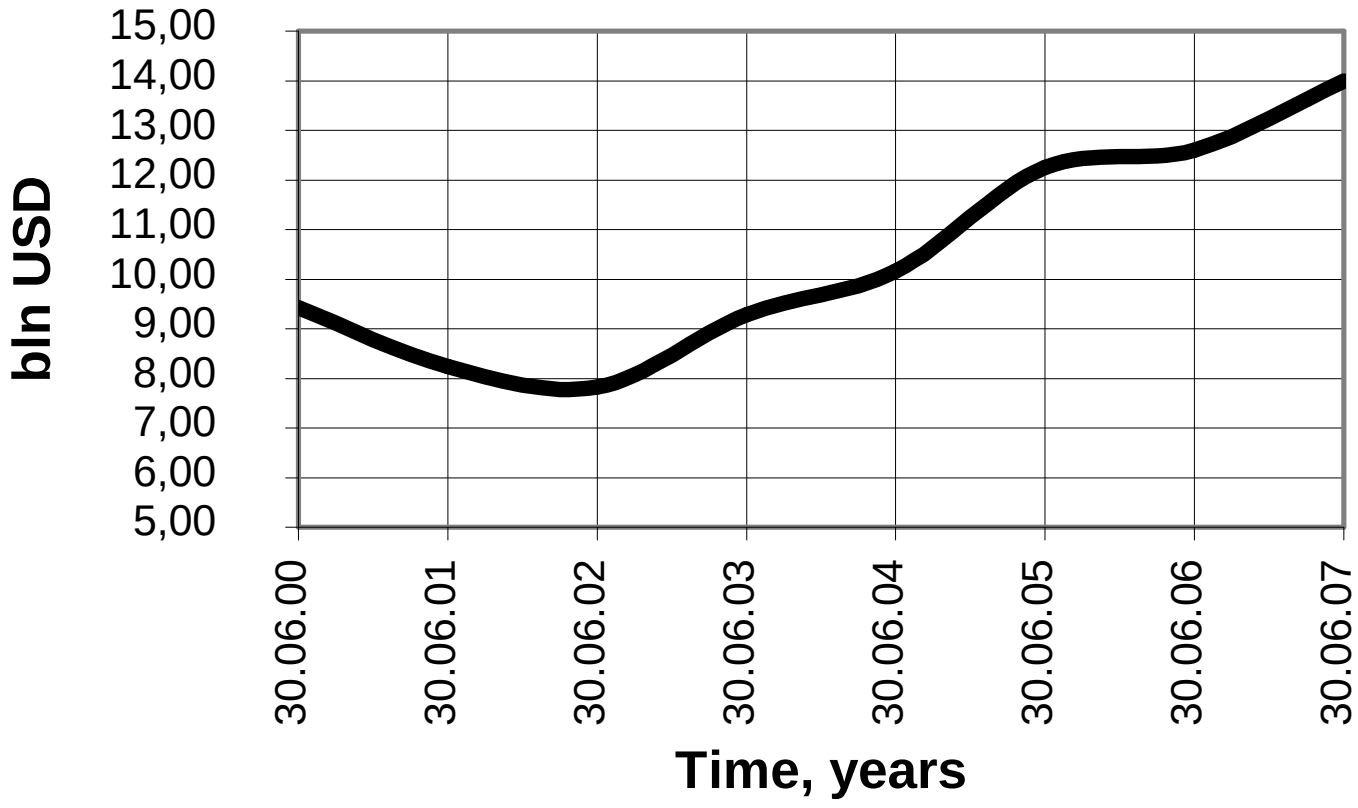

Fig.3. Trajectory of Microsoft Co. in the business space 'money-time' since 2000 to 2007 [15]

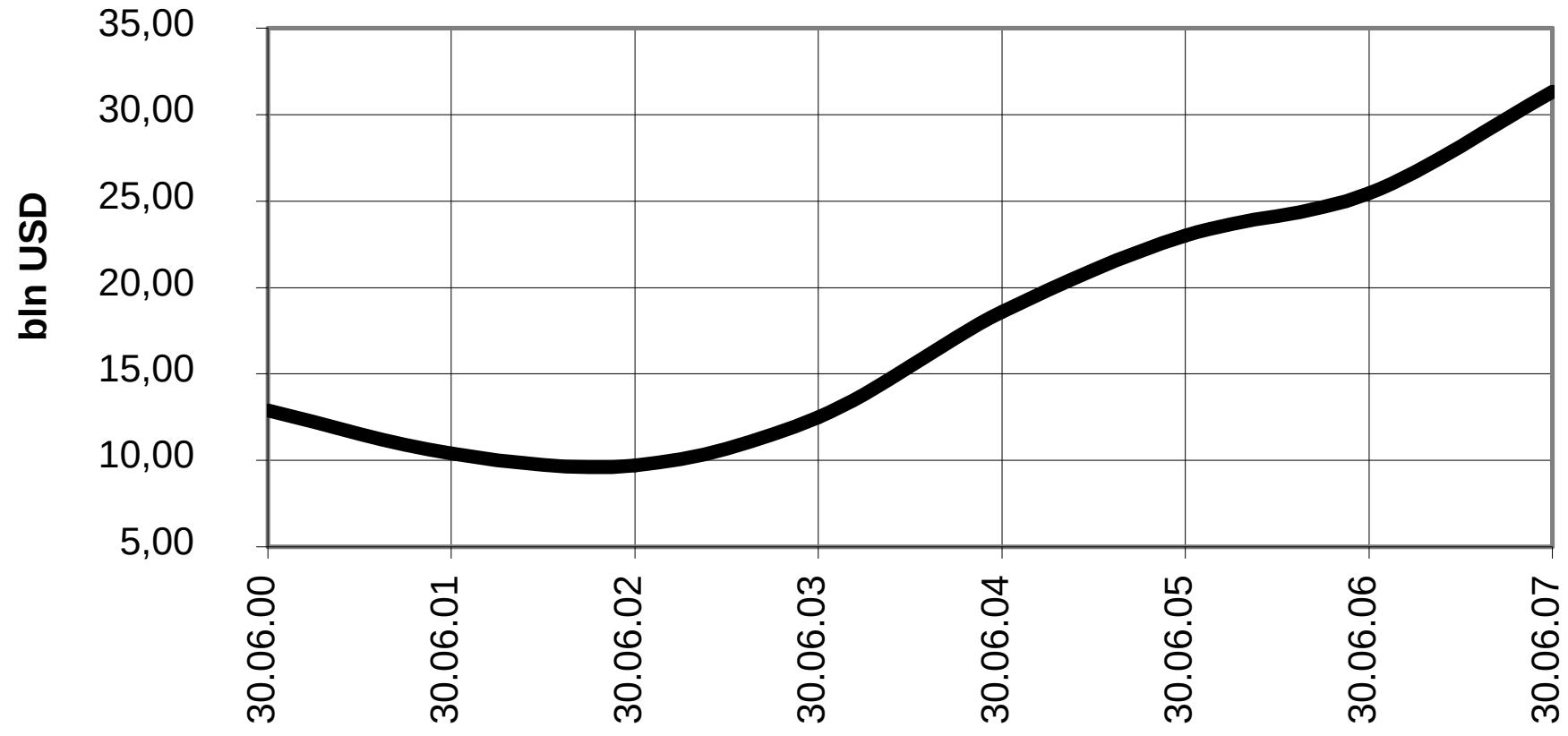

Fig.4. Trajectory of Royal Dutch Shell in the business space ‘money-time‘ since 2000 to 2007 [16]

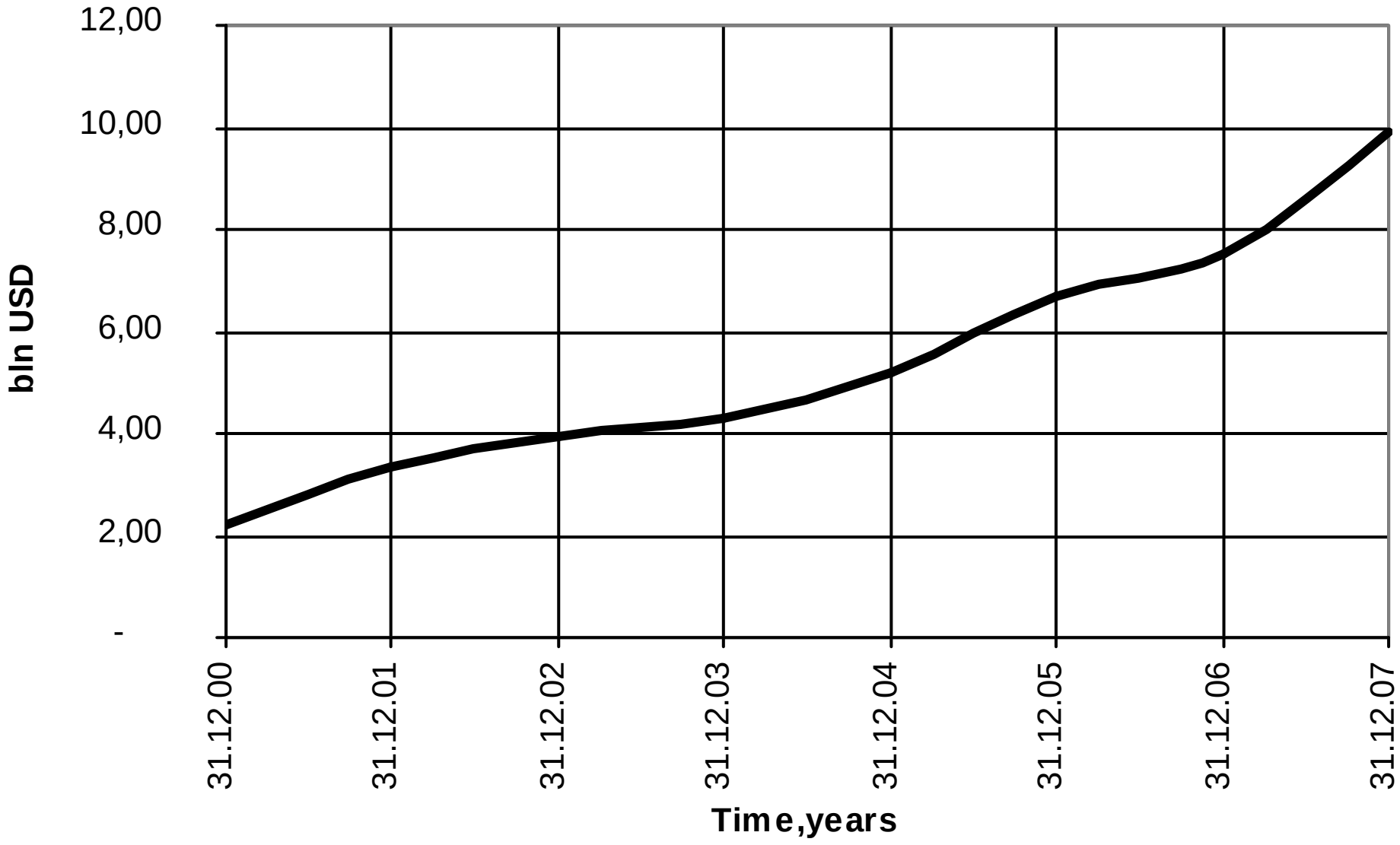

Fig.5. Trajectory of China Mobile in the business space ‘money-time‘ since 2000 to 2007 [17]

We have chosen absolutely different world companies knowingly. Microsoft is an American soft company, Royal Dutch Shell is European oil and gas company and China Mobile is the largest mobile operator in China.

From the comparison of fig. 1, 2 and fig. 3-5 we can say that trajectories of business companies very much look like the trajectories of quantum micro objects.

If we make terms shorter and leave the degree of accuracy unchanged neighboring measurements will give of course close values but the results of successive measurements will not correspond to any smooth curve and will be absolutely anyhow scattered. They particularly do not aim at being on one straight line when terms $\Delta t$ aim at null.

We must know introduce an admission that will be used later. In the previous paragraph we spoke about the free electron. But as it follows from the thoughts above motion of the company in the business space can be considered as free with great reserve because of competitors and antimonopoly law. All this facts limit business activity of the company. That is why it is reasonable to examine not free electron but measuring if its coordinates when the electron is located in potential well, i.e. its motion is restricted. For the quantitative comparison of quantum micro object and the company let us look at graphs. (fig. 6-8).

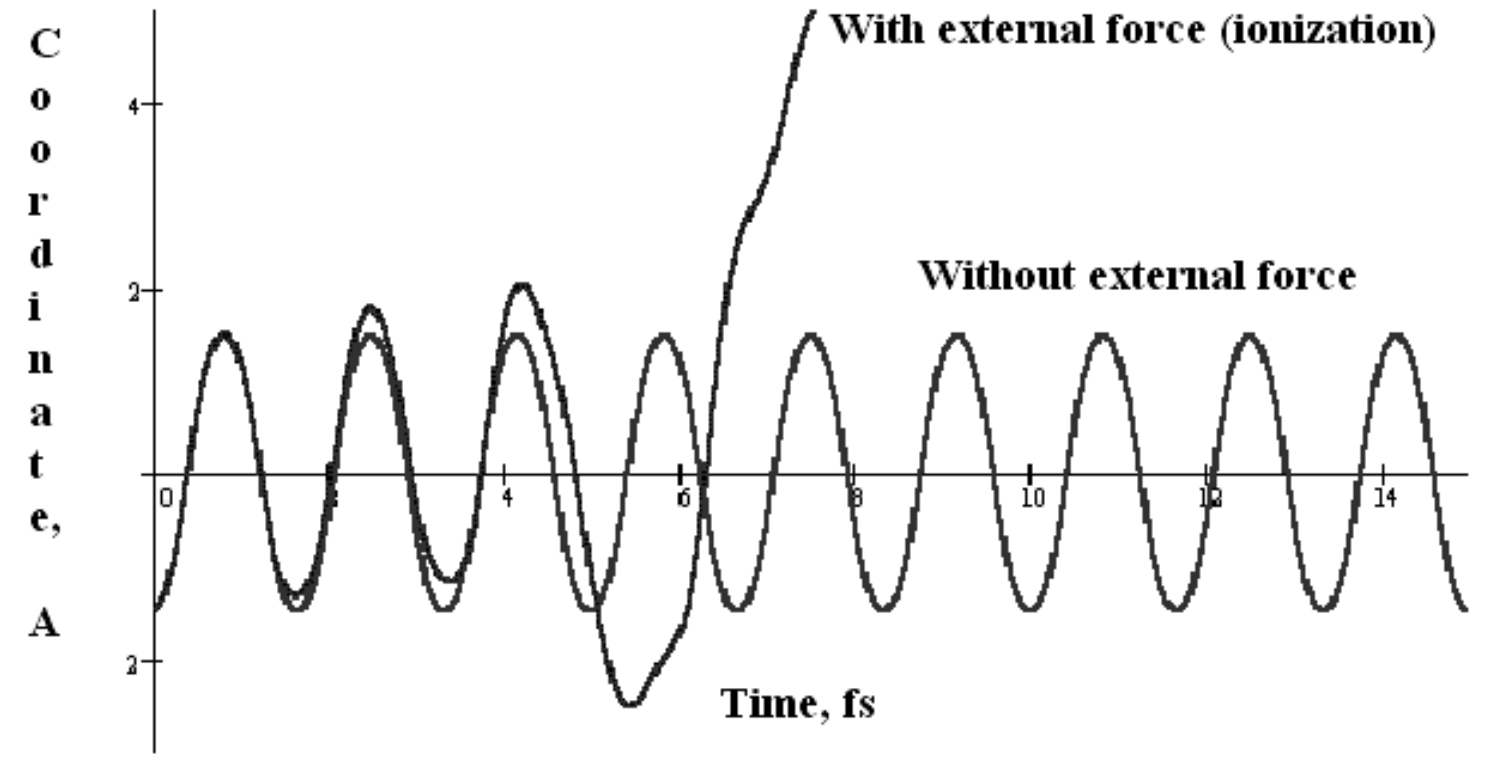

Fig. 6. Dependence of electron’s coordinate in dependence on time (measurement term is $10^{-15}$ s) [18]

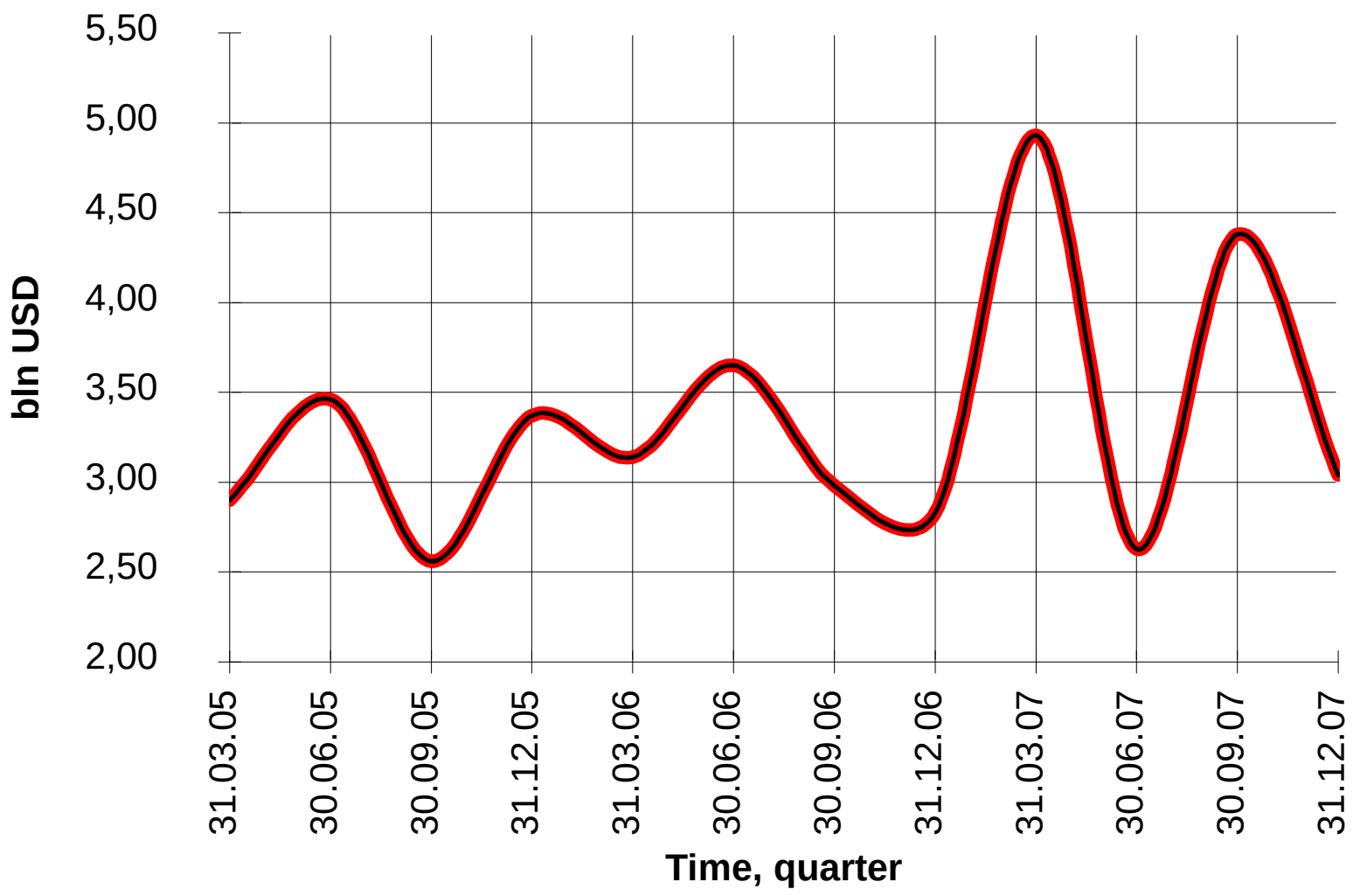

Fig. 7. Coordinate of Microsoft in dependence on time (measurement term is one quarter)[15]

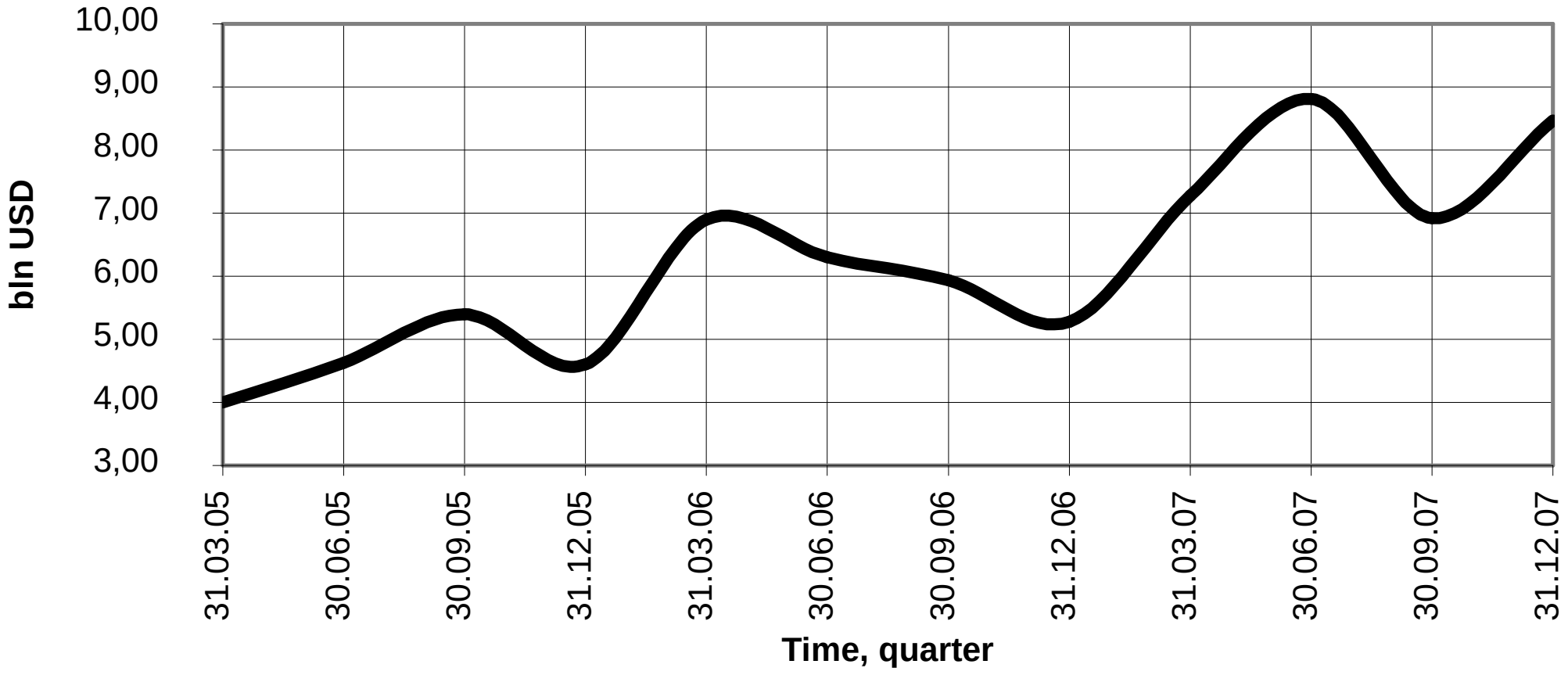

Fig. 8. Coordinate of Shell in dependence on time (measurement term is one quarter) [16]

As we see leaving degree of accuracy unchanged and decreasing of terms we receive the results for both an electron and the companies that are not located on any smooth curve (a curve with a tendency to growth or decrease). The situation is quite the opposite to the one with long terms of measurement.

So even simple half-quantitative comparison of business company and quantum micro object behavior approves the fact that the laws of their description are very similar.

One more interesting example. In the fig. 6. we can also see functional dependence of electron's coordinates on time in case of external force presence. The electron receives some supplement energy and becomes free for some time.

It is easy to note that fig. 9. is a mirror-like copy of ionization curve in fig. 6.

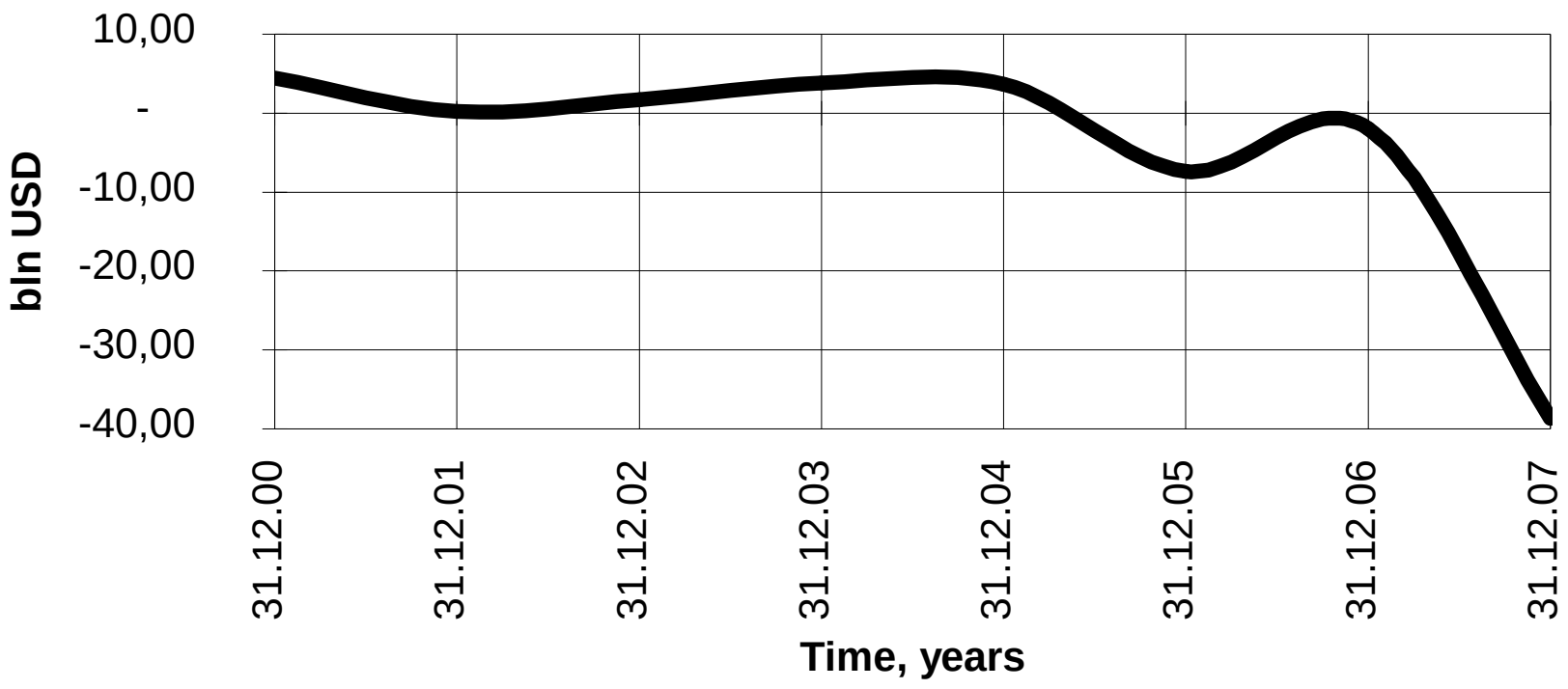

Fig.9. Trajectory of General Motors in the business space 'money-time' since 2000 to 2007 [19]

Indeed since 2005 GM began to do the business with losses aiming at conservation of leadership in the sales market. Eventually such a policy of the company led to the record losses in 2007. So we can say that GM refused from accepted rules existing in 'potential well' of automakers and having received supplement energy left the common market for a short period of time. It will probably return to it very much like a free electron becomes again bound in a very short term.

For the more accurate investigation it is necessary to define the values of effective masses of the given companies and space intervals of the events happening more precisely.

## Results discussion

So it was shown that quantum micro objects and business structures have much in common particularly trajectory absence and the discrete character of energy transfer process. As it is known that according to the hypothesis of the ideal market every man possesses full information about all players in the market. It is obvious that in a little bit complex market, but any real market is of course complex, this hypothesis does not work. We should replace it by the idea of incompleteness (imperfection) of information. Some know more and they win, the others know less and they lose. Also the model of economical equilibrium based on this hypothesis does not work. We need not static but dynamic equilibrium for prediction. We have to predict the equilibrium in a week, in a month, in five years. Such a prediction can be possible if people begin to estimate not today but future state of affairs.

For making management decisions the information about past, current and future state of the economical system is necessary every time. Also it is very important to have information about prevalent tendencies in the market and particularly about future. For making strategic decisions it is necessary to predict the development of economical situation in the country or in the world for 30 - 100 years. Long-range forecast are used for this purpose. Main demands to any forecasts are accuracy and reliability. To have a reliable forecast for 30 years we need information about the economics during quite a long period of time. [20] This circumstance makes Long-range forecasting rather complex at the stages of both collecting information and forecasting.

However the history of successful companies teaches that their best decisions were often not the results of detailed strategic plan but the results of experiments achieved by trial-and-error method and frankly speaking chances. Bill Hewlett who was one of the founders of HP company said ‘Even in the key 1960-s HP never planned further than for 2-3 years. ‘ [21] Another example: in average it takes *four* years for successful companies to elaborate strategic concepts of their development.

Frankly speaking we think that influence of generalized uncertainty relation can be distributed not only on economical processes but also on description of alive systems behavior and functioning of different mechanisms or public institutions.

Charles Darwin in his famous ‘Origin of species…‘ (1859) wrote ‘I prefer a thought that instincts [of well adapted species] are not specially granted or created but they are small consequences of one general law which determines the evolution of all alive creatures…‘ [22]

And indeed every person is a small part from the huge mass. Life of a man and his ancestry is a long chain of successive and very often accidental events. In the history of mankind chance is not the last factor. Somebody can hardly risk predicting the exact coordinate and impulse of the mankind not in a century or millennium but even in a year. One of the authors was taking his blood pressure and pulse daily for 3 months 3 times a day. We can see that the results also do not correspond to any smooth curve. (fig.10.)

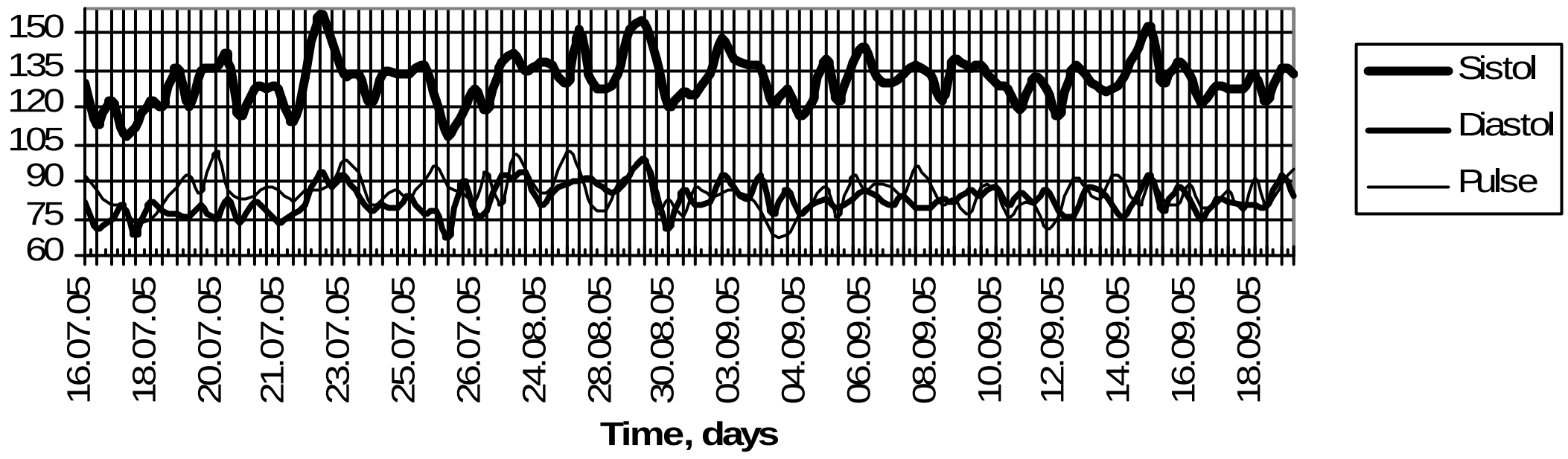

Fig. 10. Pressure and pulse dynamics of one of the authors

In conclusion let us consider one potentially successful approach of finding the numerical value of $\hbar_{gen}$ based on Bohr’s atomic theory. According to this theory every type of atoms is characterized by the succession of quantized energy values which correspond to different stationary states. Atom’s transfer from one stationary state to another is accompanied by radiation. Bohr supposed that every spectral line corresponds to instantaneous transfer of the atom from one quantum state to another which is characterized by smaller energy value. Supplement energy is emitted by radiation. In quantum theory it is quite natural to believe that the energy is emitted as separate quanta, or photons. So while the transfer an atom emits photon whose energy equals the difference of energies of final and initial states. From this we immediately have so called frequency rule by Niels Bohr. Frequency of spectral line which corresponds to the transfer of an atom from state A to state B equals the difference of energies of an atom in states A and B, divided by Planck’s constant. As for mathematical expression Bohr’s theory had one great disadvantage. Indeed it allowed finding energy of stationary states only for merely circular motion even in the simplest case of hydrogen atom. The reason was in the absence of appropriate quantization methods as Planck’s method was good only for one-dimensional motion. Therefore it was necessary to offer new quantization methods which could be used in *multivariate case*. This problem was solved in 1916 simultaneously by Wilson and Sommerfeld. They paid attention to the fact that all mechanical systems considered in quantum mechanics are *quasi-periodic systems with separable variables*. Such systems are characterized by *periodical changing of all variables although the periods are different.* Moreover with a good choice of these variables it is possible to break up the integral on a number of integrals and each of them depends on only one variable. Integrating each one by the period and equating the result with Planck's constant multiplied by integer we obviously have quantization methods for the systems with many degrees of freedom.

Are not systems considered in our work such quasi-periodic systems in *multivariate* space? If it is so we can use frequency rule and Sommerfeld quantization method for finding $\hbar_{gen}$ . Kondratiev theory of long waves is the answer to this question.

In the middle 1920-s Russian economist Nikolai Kondratiev (1892-1938) offered the theory of long waves (40-60 years) of economical market situation theory (fig.11).

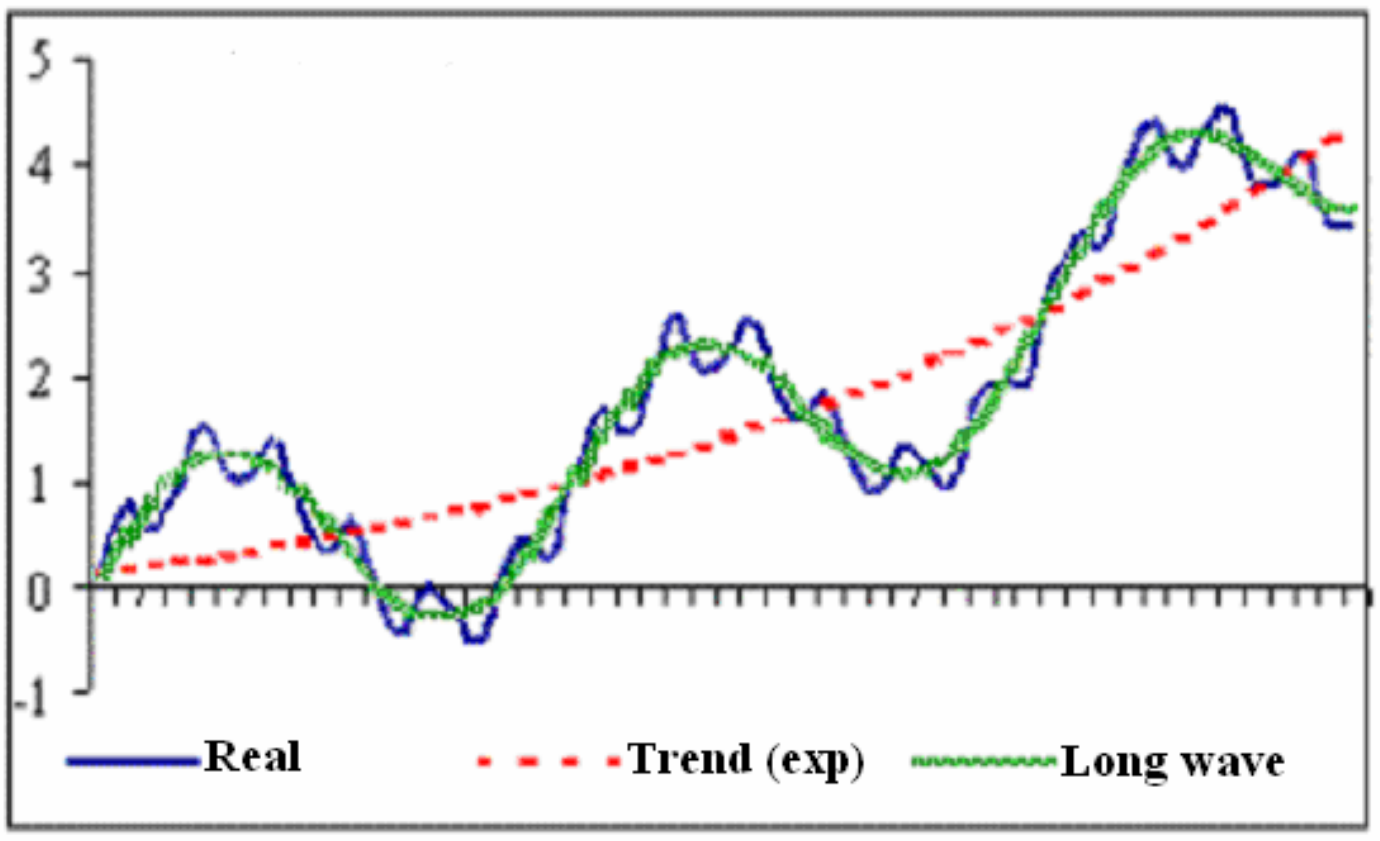

Fig.11. Elements of macro indicator's dynamics [23]

He had predecessors who managed to guess right the existence of the long wave in the economics but only after Kondratiev's fundamental work[23] serious investigations in this area began. He analyzed some economical indicators of west European countries and the USA since 1790 to 1920 (there was no more accurate statistics at that time). Having graphed the results and after smoothing of short-term changes he found out that values of these indicators synchronously move during the long wave. Maximums were reached approximately in 1815 and 1873; minimums were in 1845 and 1896. Besides on the rise of the long wave the amount of wars and revolts in the world increased and more active involving of new countries into the world trade and world division of labor happened. According to these observations Kondratiev made a long-range forecast until 2010 having predicted from example the Great Depression of 1930-s.

1 long wave - since 1779 to 1841-43 (growth phase - until 1814; reduction - since 1814 to 1841-

43).

2 long wave - since 1844-51 to 1890-96 (growth phase - until 1870-75; reduction - since 1870-75 to 1891-96).

3 long wave - since 1891-96 to 1929-33 (growth phase - until 1914, reduction - until 1929).

4 long wave - since 1929-33 (probably until the end of 1930-s) to 1973-75 (probably until 1981); acme – the middle 1950-s.

5 long wave - since 1973-75 to 2010-15; acme – the middle 1990-s.

As we see quantum economical objects satisfy fundamental laws of quantum mechanics.

It is necessary to note that the explanation of the long wave theory is still not found.

To our mind it is nothing more nor less than influence of generalized uncertainty relation.

Thus to find $\hbar_{gen}$ we have to solve a problem to some extent inverse to the problem of energy values finding. We can use the same frequency rule. We know energetic spectrum of companies (let it is annual net profit) (fig.12).

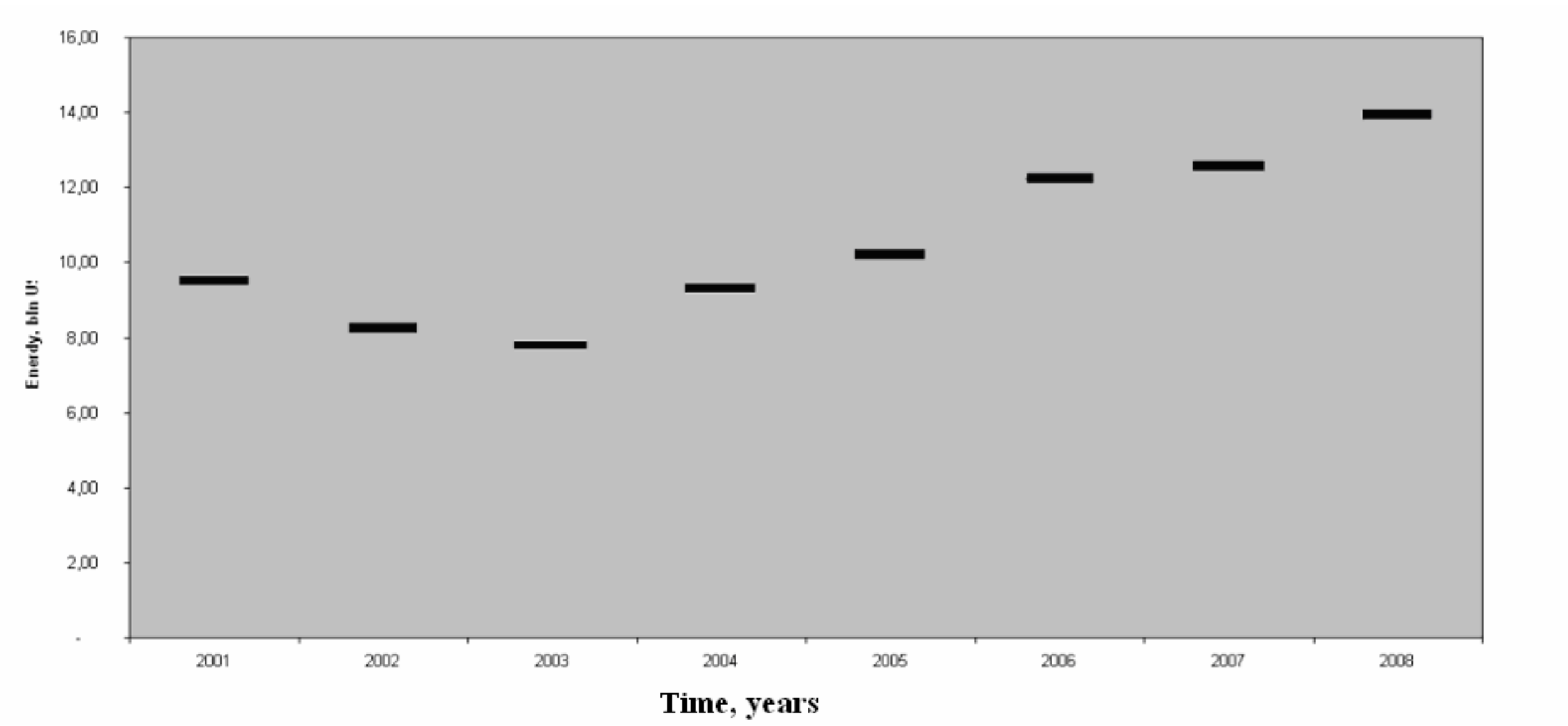

Fig.12. Energetic spectrum of Microsoft Corporation [15]

If we divide the difference of energy values by an effective frequency we will easily find $\hbar_{gen}$.

Every company like any physical quantum system has some structure which determines its energy spectrum. The size of the system also influences the structure of the system. Structure of the company is an arrangement of separate structural elements connected by official and other inner instructions and also by orders of direction and stockholders. It is a pity but now we are not able to classify all types of structures, possible types of companies' symmetry and also find the probability of quasi-structures existence (like quasi-crystals) to introduce the analogue of band theory in quantum theory of economics. But we are sure this way is the right one in searching of numerical value of an effective frequency and then $\hbar_{gen}$.

We should say that there is probably more simple way of finding $\hbar_{gen}$. The fact is that the majority of economical systems are strongly correlated i.e. they are in strongly correlated states in dependence on the environment and other participants of economical process. In other words the level of influence of these systems on each other is great. As it is known from physics for such cases the uncertainty relation is modified:

$$\delta x \delta p \geq \frac{\hbar_{gen}}{2\sqrt{1-r^2}} = \frac{\hbar_{eff}}{2}, \tag{23}$$

where r is so called correlation coefficient, $\hbar_{eff}$ is effective generalized Planck's constant. As we see for strongly correlated systems (r~1) $\hbar_{eff}$ can be much bigger than $\hbar_{gen}$. It is obvious that in dependence of the correlation level, whether the system is strongly or weakly correlated ($r \rightarrow 0$) $\hbar_{eff}$ can change in very wide range. Thus we do not need to classify all companies according to their structure but can only divide them onto 2 relative groups of strongly and weakly correlated ones. For each group using statistical data on net profit (energetic spectrum of the company), level of its correlation with the environment (value of r) we can find average trajectories of the companies (dispersion in impulse and coordinate) and then limit values of $\hbar_{gen}$. The problem is that in this case when we take the property of inseparability of quantum system into account we inevitably lose the universality of generalized uncertainty relation and each time have to find the value of $\hbar_{eff}$ while analyzing the given company.

Of course we should carefully draw analogies between business and physics. We do not think that all changes and progress of the companies are the consequences of unregulated physical process. It would be not correct to believe that companies are completely similar to quantum objects. Restating Landau and Lifshitz coursebook[7], we can say that we are speaking about the economics meaning any quantum object, i.e. an event or a system of events which obey quantum theory laws and can not be described with classical approach. At least no one from world leading economists offered such an approach. We underline once more that fact that to our mind the influence of generalized uncertainty relation is not limited by the economics only and can be distributed on the wide range of macro systems and processes in them.

In addition let us produce one interesting observation. In Russian economic mass media profit of leading Russian companies is often calculated 'by RSA' or 'by IFRS'. What is it?

RSA (Russian Standards of Accounting) is the total of forms of the federal statute of Russian Federation which regulate the rules of accounting.

IFRS (International Financial Reporting Standards) is a set of documents (standards and interpretations) which regulate the rules of financial reporting compilation which is necessary for making economical decisions by external users. In the majority of european countries financial reporting of the companies whose shares are sold on the exchange must be compilated 'by IFRS'. One of the principle differences of RSA from IFRS is the hard regulation of accountant work. That is why Russian accountants who have not got used to relative freedom in their work meet significant difficulties during transformation of reporting according to IFRS.

RSA traditionally head for inquiries of regulating organs first of all tax ones whereas IFRS head for users who have real or potential interest in the subject of reporting namely stockholders, investors and contractors.

RSA does not suppose the consolidation of reporting for holding companies and this makes the analysis essentially more difficult since only the activity of the head company is reflected in the report without the activity of subsidiaries.

IFRS are the standards based not on rigid requirements but on principles. The aim of IFRS is an arrangement of such conditions for the users in which they can follow principles and do not need to look for loopholes in the rules.

Since 1998 the reform of accounting according to IFRS has been fulfilled in Russia. Since 2005 all credit organizations in particular have to prepare their reporting according to the norms of IFRS.

To our mind the beginning of Russian system of accounting reform according to IFRS is a good proof of that fact that conversion of Russian economy to market state nevertheless happens. But moreover choice of IFRS as prevalent system of financial reporting in the majority of countries with market economy approves once more all said above in our work because IFRS include taking principles of quantum uncertainty into account as it follows from the properties of IFRS. And as it was shown in our work before quantum uncertainty is a natural property of market economy. Using of RSA in our country is echo of the past. It is well known that planned economy of the USSR did not allow any quantum uncertainty in financial indicators as all factories of Soviet economy worked according to the plan which was elaborated for several years ahead.

Despite the fact that RSA is hardly regulated there are often legal action of tax departments against some companies. In some cases these actions are direct evidences of deterministic approach to quantum objects uselessness. In other words determinism does not allow in principle to describe all possible situations that can happen with such an object as a company in frames of that quantum space where it exists.

We should make a reservation that construction of quantum theory of economics is impossible without using of classical economics principles as instruments for 'measuring'. As in quantum mechanics the task of quantum theory of economics could be the definition of the probability of the ‘measurement‘ result.

Possible enemies of such a quantum approach to the description of macro structures will be probably interested in the experiment carried out by American scientists from University of California, Santa Barbara [24]. They made rather sensible device namely vibrating crystal bar more than one micron lengthwise and ultrasensitive detector which allows registering displacement of the bar on distances about thousandth of nanometer in length. The bar consists of approximately 10 milliard atoms so in comparison with an isolated atom the bar can be considered as macro object. If they managed to demonstrate the fulfillment of uncertainty relation for the bar this would become the bright example of quantum phenomena manifestation outside the realm of microscopic physics.

Let us consider the following example to understand the main idea of the experiment. If we press one end of wooden line to the edge of the table and then pull by another end the line will begin to vibrate with dying oscillation and come to quiescent state in some time. However if we look at the free end of the line through a powerful microscope we will see that it shakes disorderly by quiver. This shaking is the consequence of accidental strikes of air molecules on the end of the line and also existence of many fluctuating inner defects. But there are also quantum fluctuations in the line which can not be seen because of classical heat motion. This ‘zero‘ fluctuations are much less in the amplitude and appear because of quantum uncertainty of coordinates and impulses. By analogy with hydrogen atom elastic back moving force which affects on the line is balanced by the repulsive action of fluctuating velocity of the line’s center of gravity. The crystal bar played the pole of the line in the experiment.

Since ‘zero‘ fluctuations of coordinate and velocity are extremely small they can be observed only after suppression of usual heat fluctuations. So it is necessary to cool the system up to very low temperatures. In the structure with frequency of characteristic mechanical oscillation ~ 1GHz (this frequency corresponds to the frequency of bar oscillation in the experiment) ‘zero‘ fluctuations become prevalent when the temperatures become ~ $10^{-2}$K. Authors estimation showed that the uncertainty of the bar displacements in this case is ~ $10^{-5}$nm. They used single-electron transistor for their registration. The transistor was a tiny metal ‘island‘ compared in its sizes with the bar and separated by dielectric barriers from two wires from voltage source. The barriers were thin enough for the electrons could tunnel from one wire to another through the ‘island‘. The voltage was chosen in that way that only one electron could tunnel through the ‘island‘ for one time. Such a single-electron current is very sensitive to fluctuations of electric charge not far from the ‘island‘ what was used for registration of bar displacements. The bar was separated from the ‘island‘ by vacuum gap of 250 nm in width. Bar vibrations led to the change in gap width and corresponding redistribution of the charge on the 'island' what in its turn led to fluctuations of tunneling current and allowed to judge about vibrations' amplitudes. (fig.13.)

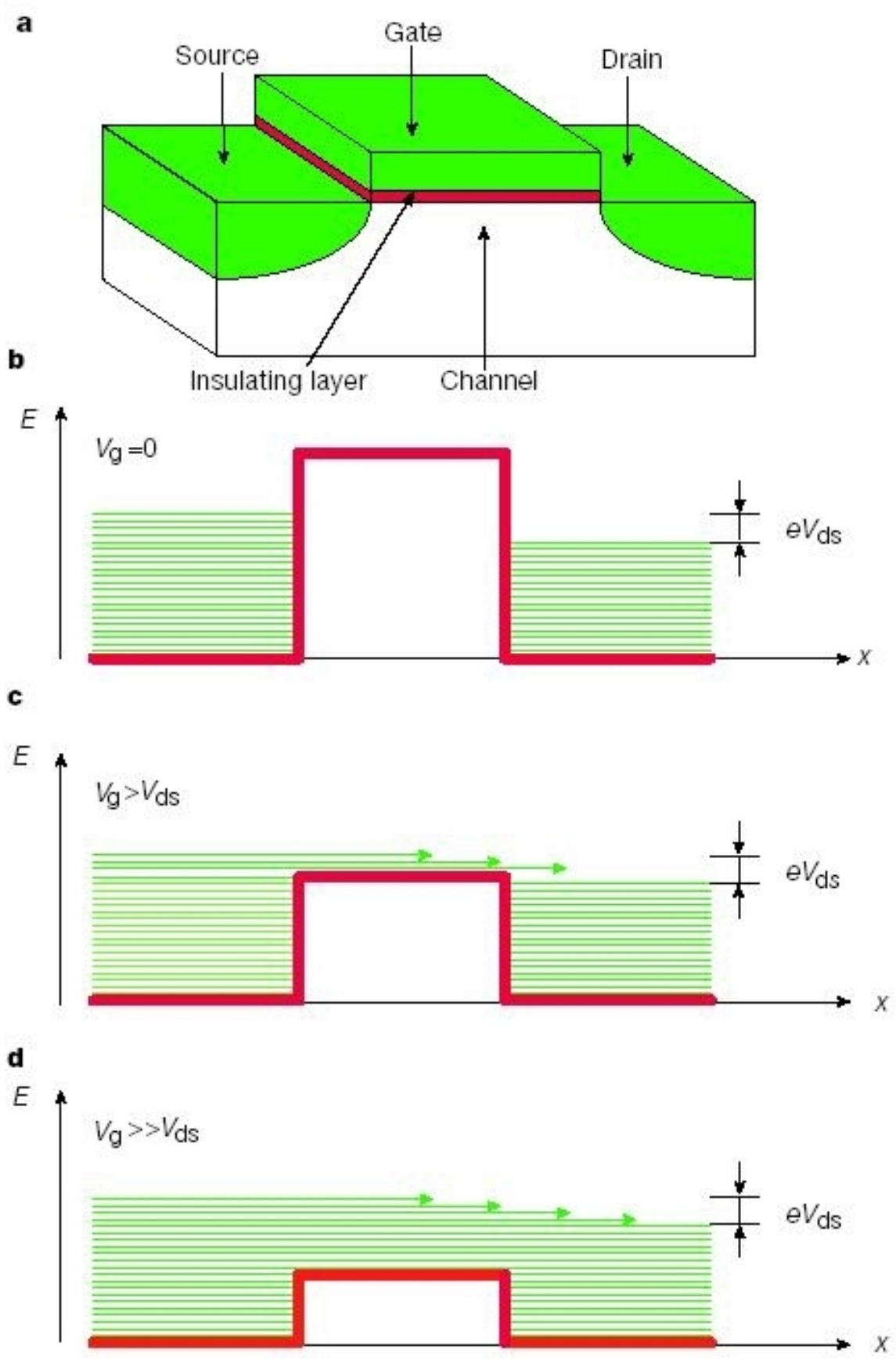

Fig.13. Principle scheme of the experiment [25]

It is a pity but the scientists were not lucky enough to reach the sensitivity for registration of 'zero' fluctuations. Thus uncertainty relation has remained unchecked for macro objects. But the progress is huge. It is necessary to multiply the sensitivity to displacements by ~ 100 and multiply the frequency of the bar by ~ 10. Taking the fast development of nano-technologies into consideration we are sure that this problem will be sold in the nearest future.

Any economist having read this article to the end can ask 'Well, and what? What can I have from the fact that an object is quantum?' Probably not so much at the present level of our theory development. However some basic facts from our work are very important.

First of all do not make vast plans on many years ahead. Do not believe analysts and forecasts because they are not able to avoid quantum indeterminacy and predict what is going to be with a company tomorrow. At least they can't do this until economical Planck's constant is determined. Analysts can offer you very beautiful differential and integral equations to describe the future of your company but they can be solved only with strictly fixed boundary conditions. And these conditions can't be formulated because of inseparability of the company from the environment. It is impossible to predict the precise trajectory of the company. We can only build an average trajectory using known from the history of the company points and try to continue this curve in future a little bit. Every chief can do this. According to Collins great companies have already understood it and make only 2-3 years ahead forecasts.

Second, as the executive officers as a rule want to decrease the indeterminacy in financial coordinates at the expense of getting things put in order and more control over employees. It is impossible to determine simultaneously the coordinates of the company in the business space with dimension N and the direction of impulse (or its velocity and necessary effective mass). For example how can we decrease the indeterminacy in coordinates of the company from the quantum mechanics' point of view? Would you like to know the coordinates more precisely in a year? Then you will have difficulties with necessary speed of grow or effective mass search. Would like to find out the necessary speed of grow of the company or the whole economy? Be ready to the huge grow of the indeterminacy in its future coordinates!

Third, try always to stay in quantum state. The company can't work properly being separated from the market. In opposite it looks like a deaf and blind man without a guide in a thoroughfare. A company with big effective mass does not feel the energetic spectrum anymore (because of the spectrum change) and continues moving itself by inertia. It is practically impossible to stop or change the trajectory of such a company. Very much like the freight train. We agree that classical objects are the objects with predictable trajectory but this way is a way to abyss or state ownership.

Fourth, it may seem rather contradictory to all said above but try to organize your business to have the minimum correlation radius r. In this case the dispersion in impulses and coordinates of the company can be much less. In other words try to be independent on the processes around the company as much as possible. But null values of r are hardly achieved in the real life.

'Companies-aristocrats' (according to Adizes term) are partly right when they hope on external situation. Life experience has taught them that the trajectory of the company can make a sudden turn another day. The fact is that the situations greatly depends on the energy and the effective mass. On other words, the companies with great effective mass can safe their trajectories rather long because of great inertia. However because of energy spending (there is no motion without friction in the real life) all kinetic energy store of the company will be spent and the impulse will be null. Nonzero rest mass will of course provide the company with some store of inner energy but the company will need essential energy spending to begin its motion again.

By the way the notion of 'overheated economy' is well to known to all economists. This term well coincides with econothermodynamic approach. [26] Velocity increase leads to the quadratic increase of kinetic energy and as a result to the increase of econothermodynamic temperature T. It is the reason why the economy is overheated.

# Conclusion

In the present work it is shown that both quantum micro objects and business companies with small effective masses have the following common features. They are inseparability, the absence of trajectory and as a result big problems with any forecasting of their behavior, small effective masses, a discreet process of interactions and energy transfer. We believe that the generalized uncertainty relation got in our paper can be used not only in case of economical processes but also for the description of other macro systems behavior. The basic reason for this is quantum nature of the space around us.

Per se our work offers the following. As confines of application are still not known let us offer the world leading economists get acquainted with works of W. Heisenberg, E. Schrödinger , Max Planck etc. and together try to describe economical processes or any other events with small effective mass on small space intervals using quantum mechanics. E. Stanley report at recent March APS meeting, regarding possibility of application of spin-glass model to describe a stock market fluctuation [27] is a solid step in the right direction because the spin-glasses are one of classic quantum objects.

However finding of numerical value of $\hbar_{gen}$ remains the central problem of quantum theory of economics. Finding this value will allow to introduce the analogue of Schrödinger equation and the elements of quasi-classical approach of Wentzel-Cramers-Brilluen (WKB method) in our theory.

Authors would like to thank K.V. Nechaev and Cand. Sc. Y.I. Spichkin for instructive discussions whiles this work preparing.

## References

1. http://worldcrisis.ru/crisis/87897
2. Niels Bohr Collected Works 13-Volume Limited Edition Set, General Editor, Finn Aaserud; ISBN 978-0-444-53286-2 **Volume 12.** Popularization and People (1911-1962)
3. Bohr, N., Atomic Physics and Human Knowledge (1958), Wiley Interscience, 1987 Ox Bow Press: ISBN 0-91802452-8, seven essays written from 1933 to 1957
4. E. M. Gershenson. Single atoms investigation. Soros educational magazine, № 1, 1995. p. 116-123
5. Bohm, D., 1951. Quantum Theory, New York: Prentice Hall. 1989 reprint, New York: Dover, ISBN 0-486-65969-0
6. Adizes I. Managing corporate lifecycles. - Paramus, N.J.: Prentice Hall, 1999.- 460.
7. Landau and Lifshitz Course of Theoretical Physics vol. 3: "Quantum Mechanics: Non-Relativistic Theory". L. D. Landau, E. M. Lifshitz 2001
8. Titus Lucretius Carus *On the Nature of Things*, (1951 verse translation by R. E. Latham), introduction and notes by John Godwin, Penguin revised edition 1994, ISBN 0-14-044610-9
9. http://www.mirrabot.com/work/work_68265.html
10. Built to Last: Successful Habits of Visionary Companies - Jim Collins, Jerry I. Porras. Harperaudio. November 2004.ISBN-13: 9780060589059

11. http://rating.rbc.ru/article.shtml?2004/08/30/740555
12. Einstein, Albert (1969), Albert Einstein, Hedwig und Max Born: Briefwechsel 1916–1955, Munich: Nymphenburger Verlagshandlung
13. Physical encyclopedia, vol.3, Moscow, ‘Big Russian encyclopedia’.1992
14. http://040.help-rus-student.ru/pictures_fail/
15. http://www.microsoft.com
16. http://www.shell.uz
17. http://www.chinamobileltd.com/
18. http://www.1580.ru/album/2001/26-01-01/index.html
19. http://www.gm.com/
20. The long Wave Debate. Selected papers from an IIASA International meeting on Long-Term fluctuations in economic growth: their causes and consequences, Held in Weimar, GDR, June 10-14, 1985.
21. Good to Great: Why Some Companies Make the Leap... and Others Don't. Jim Collins. Collins. 2001. ISBN: 0066620996
22. On the Origin of Species by Means of Natural Selection, or the Preservation of Favoured Races in the Struggle for Life, - Charles Darwin. 1859. Print (***Hardback*** & ***Paperback***). ISBN 0-486-45006
23. Kondratieff, Nikolai D., "The Long Waves in Economic Life," Review of Economic Statistics. 17(6) Nov 1935
24. R.G.Knobel and A.N.Cleland, Nature. 2003, 424, 291
25. M.H.Devoret and R.J.Schoelkopf, Nature. 2000, 406, 1039
26. A.M. Tishin, O.B. Baklitskaya EconoThermodynamics, or the world economy "thermal death" paradox, arXiv:0807.0372 , 27 pages, http://lanl.arxiv.org/abs/0807.0372
27. APS News, May 2008, p.3